\documentclass[twoside,letterpaper]{article}

\usepackage{graphicx,epsfig}
\usepackage{amssymb,amsmath, amsthm,url}

\newcommand{\ven}{{\vec{\eta}}}

\title{    Electro Thermal Transport Coefficients at Finite Frequencies }
\author{ B Sriram Shastry \\
Physics Department, University of California,  Santa Cruz, California 95064 }
\date{\today}

\begin{document}
\maketitle

\begin{abstract}

We review a recently developed formalism for computing thermoelectric  coefficients in correlated matter. The usual difficulties of such a calculation are
circumvented by a careful generalization  the transport formalism to finite frequencies, from which one can extract the high frequency objects. 
The technical  parallel between the Hall constant and the Seebeck coefficient is explored and used to advantage. For small clusters, exact diagonalization 
gives the full spectrum for the Hubbard and especially the  $t$-$J$ model, a prototypical model for strong correlations, and this spectrum can be used to compute the exact finite frequency  transport coefficients and hence to benchmark various approximations. 

An  application of this formalism to the physically important case of sodium cobaltate $Na_xCoO_2$ is made, and interesting  predictions for new materials 
are highlighted.
 \end{abstract}

\tableofcontents

\section{ Introduction}
\subsection{The Challenge of  Correlated Electron Systems.}

Correlated electron systems\cite{lectures,reviews_1,reviews_2,reviews_3,reviews_4,lee,fermi_liquid_theory_anomalies_1} stand at the frontier of Condensed Matter physics, posing conceptual as well as calculational hurdles that have seriously engaged the theoretical  community in the last few decades. Experimental results on several classes of new materials, have provided a great impetus to this study, and often given direction to the theoretical endeavors. High $T_c$ superconductors is a large class of materials that are within the domain of correlated electron systems, but not the only ones. The study of rare earth compounds provides another important class of systems, as do the newly discovered cobalt oxide materials.

Correlated electron systems are characterized by a common feature, namely a narrow bandwidth of electrons,  which interact strongly  on a scale of the order of electron volts at short distances. A dimensionless coupling constant, namely the ratio of the interaction energy U($\sim$ a few eV) to the band width W( $\sim$ .1 or 1 eV) becomes  large. This large  parameter makes the validity of a perturbation theory in $U/W$  unclear. In weakly interacting systems, such as good metals, the analogous ratio is small, and leads to the Fermi Liquid picture of weakly interacting quasiparticles, as formulated by Landau and others. In the case of correlated matter, as in many other settings, the behaviour of  perturbation theory  has a strong dependence on spatial dimensionality.  In 1-dimension,  the standard  Fermi liquid theory breaks down due to a proliferation of low energy excitations, i.e. an infrared breakdown. The most interesting case of 2-dimensions, i.e. electrons moving in a plane, is the hardest problem yet, since special techniques that work in 1-d are not applicable here, and yet the low dimensionality suggests enhanced quantum  fluctuations. This case is of experimental consequence, since many correlated materials are also layered, displaying a large asymmetry between their transport properties along  planes and across these planes.

The basic models that have been used to describe  correlated electrons are the Hubbard and $t$-$J$ models described below, and the periodic Anderson and Kondo lattice models. More complex models with multiple bands have been considered, but in this article we shall restrict our attention to the first two models, which describe a very large class of systems where d-type electrons are involved.
In physical terms the Hubbard model $H= H_0 + H_1 $ contains the hopping of electrons between  sites denoted as  $H_0 = -   \sum_{\ven,\vec{x}} t({\ven}) \; c^\dagger_{\vec{x} + \ven, \sigma } c_{\vec{x} \sigma }$, where $t(\ven)$ is the   hopping matrix element for a range vector $\ven$ \footnote{ The band width on simple lattices is related to the hopping through  $W\sim  2 \nu |t|$ where $\nu $ is the coordination number of the lattice. The vector $\ven$ is usually, but not always, specialized to be the set of  nearest neighbour vectors on the lattice. },  and an on-site Coulomb repulsion terms $H_1 = U \sum n_{j \uparrow }n_{j \downarrow }$. This model  neglects all other (smaller) terms in the full lattice Coulomb problem.  It is thus the simplest correlated electron model, characterized by the dimensionless coupling $U/|t|$  and the filling of electrons in the band denoted by $n = N/N_{s}$ (where N and $N_s$ are the total number of electrons and the number of lattice sites), so that from the Pauli principle we are restricted to the range $0 \leq n \leq 2$. The so called ``Mott Hubbard gap'' arises in this model at half filling $n=1$ as follows. At this filling, there is a single electron per site on average, and so  it is impossible to avoid paying an energy penalty of O(U) on adding a particle, but it is quite cheap (independent of U) to remove a particle. Thus the cost of adding a particle is quite different from removing a particle.  This fundamental  asymmetry characterizes an insulating state in the most  general possible  terms. It does not  invoke any kind of  broken symmetry whatsoever. Such an insulator is called the Mott Hubbard insulator. The standard example of this kind of an insulator is    the 1-d Hubbard model at half filling.

\begin{table}[t]
\caption{{\bf  Common symbols used and their meaning}} 
\begin{tabular}{|c|c|} \hline
$t(\ven) $ & Hopping matrix element for a distance $\ven$\\
$<i,j>$& Nearest neighbour sites $R_i, R_j$ \\
$c_{j,\sigma}$ & Electron destruction operator at site $R_i$ for spin $\sigma$  \\
$N_s, N, n $& Number of sites,  electrons and density\\
$v, \Omega$& Volume per cell and total volume \\
$q_e, c $ & Carrier charge and the velocity of light\\
$U, J$ & Interaction coupling constants in the Hubbard and $t$-$J$ models\\
$P_G$ & Gutzwiller projection operator removing doubly occupied sites\\
$\mu, \rho_0(\mu)$& Chemical potential and density of states per spin at that energy\\
$K=H- \mu \hat{N}$ & Grand canonical Hamiltonian\\
$C(T), T $ & Specific heat and temperature $T$ \\
$\omega_c= \omega+ i 0^+$ & Frequency with small imaginary part\\
$R_H,R_H^*$ & Hall constant and high frequency Hall constant\\
$S, L, Z $& Seebeck coefficient, Lorentz number and Figure of merit\\
$L_{ij}$ & Onsager response coefficients \\
$M_1, M_2 $ & Response of charge  and energy currents to external input power\\
$ N_1, N_2$ & Response of charge  and energy densities to external input power\\
$\kappa, \kappa_{zc}$& The nominal thermal conductivity($L_{22}$)\\
&  and zero current thermal conductivity\\
$ E_x, B$ &  Electric and magnetic fields \\
$\sigma_{\alpha,\beta}$& Electrical conductivity tensor\\
$\hat{J}_x, \hat{J}^Q_x$ & Charge and heat current operators along x axis\\
$\tau^{XX}$& Stress tensor or effective mass tensor\\
$\Phi^{xx}, \Theta^{xx} $& Thermoelectric and thermal  operators\\
$\psi(\vec{x}),\phi(\vec{x})$ & Luttinger's gravitational field and electric potential\\
$K(q), \rho(q)$ & Heat and charge densities\\
$D_Q,D_c$ & Heat and charge diffusion constants\\
$\chi_{A,B}$ & Susceptibility of operators $A,B$\\   
$\epsilon_n$ & Energy levels of the Hamiltonian \\
$\vec{S}_j$ & Spin vector at site $R_j$\\
\hline
\end{tabular}
\end{table}

Another important description of correlated electron systems  is  through the $t$-$J$ model . This model represents a  much stronger version of correlations, with the prohibition of double occupancy, i.e. states with $n_{j \uparrow}n_{j \downarrow}=1$. This constraint is  enforced by the  Gutzwiller projection operator 
\begin{equation}P_G= \prod_j(1- n_{j \uparrow}n_{j \downarrow}),\label{gutzwiller}
\end{equation}
 so that $H_{t-J}= P_G T P_G + \mbox{exchange}$. This situation corresponds to taking  $U\rightarrow\infty$ in the Hubbard model. The added exchange term is written as $J\sum_{<i,j>}\vec{S}_i.\vec{S}_j$, and represents the tendency, arising from eliminating high energy states with double occupancy,  of neighbouring spins to point in antiparallel directions\footnote{By performing degenerate perturbation theory at large $U$ one can obtain the $t$-$J$ model and another small three body term that is neglected here\cite{hubbard_tj}.}.

Given these simple looking models, the task is to compute physically measurable variables, such as the thermodynamic response functions, as well as the dynamical response functions.  One wants to understand the nature of the order in the ground states that arise, and the dependence of these on various parameters of the models.
Various calculations suggest highly nontrivial magnetic and superconducting states to emerge from these simple models. Among the dynamical aspects of the problems,  transport variables such as the resistivity, form the bulk of measurements carried out in laboratories; these are the  table top experiments of condensed matter physics. Here one applies an electrical and a  thermal gradient along the sample (say along the x axis), and in some instances a uniform magnetic field transverse to the electric field (along the  z axis). The measured objects are the electrical and thermal currents that are induced, and by taking the ratio of currents to fields, one deduces the various conductivities $\sigma_{\alpha,\beta}=\partial J_\alpha / \partial E_\beta$, and from these   the resistivities (see Eq(\ref{hall},\ref{onsager1},\ref{onsager2}) below).
They  are relatively easy to measure, and reveal the characteristics of a given material to a very large extent, e.g. whether it is a metal or an insulator, the carrier charge and  density etc.  Next to  resistivity, one of the most widely used measurements is that of the Hall constant $R_H$  (defined below in Eq(\ref{hall})), followed by  the thermoelectric response functions such as the thermal conductivity $\kappa$  (defined below in Eq(\ref{thermal_zero_current})) , the Seebeck coefficient S and the Lorentz number L (defined below in (Eq(\ref{variables}))). The materials community has also  great interest in seeking the conditions for an enhanced Seebeck coefficient and the figure of merit $Z T$ Eq(\ref{variables}), since the overall efficiency of a device turns out to depend upon this dimensionless number. 

Unfortunately it is not a simple matter to compute these response functions. In particular these  are much harder than equal time correlations.  One needs to know more than the ground state in order  to determine these, a handle on the excitations is required as well. 
Thus  the question of  {\em transport in correlated matter},  is one of the hardest problems in condensed matter physics. The traditional  methods and models of transport  such as the  Drude Sommerfeld \cite{ashcroft_mermin,mahan} and Bloch Boltzmann\cite{sommerfield,bloch} theory  have been used extensively  to estimate answers, even when the validity of the methods is rather ambiguous in these systems.  The simplest of these is the Drude-Sommerfeld model \cite{ashcroft_mermin}, using a free electron gas within a Boltzmann equation approach,  and thereby builds in the Fermi statistics into the classical Drude theory.
 The Bloch Boltzmann\cite{sommerfield,bloch,wilson}  theory  improves on this  by focusing on  carriers in the Bloch bands, and thus the carriers  are characterized by a band index as well as a wave-vector index, and relies principally on the concept of metals and insulators as defined by band filling. This  band filling concept is often denoted as the Bloch-Wilson\cite{bloch,wilson} classification of metals and insulators.  The Landau theory of Fermi liquids\cite{fermi_liquid_theory} further refines the theory and incorporates the effect of Coulomb interactions via  renormalized quasiparticles. Thus the physical picture behind the current understanding of transport is one of   almost free quasiparticles  that diffuse through  matter, suffering multiple collisions either mutually or with the lattice and other excitations. 
 
 The above picture is not robust against strong correlations.  Mott Hubbard interactions  change the nature of the carriers radically near half filling, i.e. a single electron per atom. The Mott insulating state\cite{mott_insulator} arises as an exception to the Bloch-Wilson\cite{bloch,wilson} classification of metals and insulators. At half filling a Mott Hubbard system {\em is an insulator due to correlations}, and would have been a metal without interactions.    The strongly  correlated systems  addressed in this article  may be described as doped Mott insulators\cite{lee}, i.e. states obtained by adding or removing electrons from a Mott insulator. Here the definition of a quasiparticle has been argued to be ambiguous\cite{pwa_early,lee}, thus making the standard approach questionable. The applicability  of the Fermi liquid concept  has been questioned in strongly correlated matter,  on the basis of several experimentally anomalous results for the resistivity and the photoemission\cite{fermi_liquid_theory_anomalies_1,fermi_liquid_theory_anomalies_2}.

 In this article, we address this question from a fresh point of view, starting from the exact but usually intractable linear response formulas, generally known as Kubo formulas, and finding easier but non trivial versions of these. These new formulas approximate certain aspects of the problem that are possibly less controversial, but treat the effects of correlations  carefully.
Our results may be classified as being complementary to the usual Bloch Boltzmann  theory, and we present the formalism as well as its applications in the context of the thermoelectric response functions. In short, our method enables us to compute a well defined subset of the transport response functions, such as the Seebeck coefficient, the Hall constant and the Lorentz number as well the thermoelectric figure of merit.  This subset, described in greater detail below,  is characterized by the fact that  they are independent of the relaxation times, within the simplest Bloch Boltzmann  theory.

\subsection{Transport in Correlated Electron Systems }

As explained above, a major problem is to understand transport phenomena in correlated matter.  Traditional approaches such as the Boltzmann equation have served  long and distinguished tenures to explain transport coefficients in terms of a few measurable objects (relaxation times, and effective masses etc). However  these methods run into severe problems in the most interesting and important problem of metals, with strong Mott Hubbard correlations. These correlations give a Mott insulating state at commensurate (half) filling, with localized spins interacting with each other, and away from half filling, one has metallic states that carry the distinguishing marks and signatures of the parent Mott insulator.  

The High $T_c$ systems provide one outstanding set of materials that have dominated the community for the last 20 years. These are widely believed to be strongly correlated, following Anderson's original and early identification of these as doped Mott insulators\cite{pwa_early}. Another important material, sodium cobaltate $Na_xCoO_2$  has recently been popular in studies of thermoelectricity \cite{terasaki,ong}, this is strongly correlated too, but the underlying lattice is triangular rather than square. These two systems have in common the presence of spin half entities, and have  both been modeled in (rather extensive) literature by some variants of the $t$-$J$ model. 

The  qualitative reason  for the difficulties of the Bloch Boltzmann equation approach in these correlated models can be understood in several ways. One is to recognize from a variety of experiments   in these systems, that the wave function renormalization, or quasiparticle residue $z_k$ (defined as the jump of the ground state  occupation number at the Fermi surface in momentum space) is either zero, or if non zero, it is certainly very small.  Another simple and yet powerful  point of view is to ask: what is the  charge carrier in a Mott Hubbard system near half filling. From the real space point of view, in order that   a correlated electron can hop to a nearby position, it must make sure that there is  no particle of either spin at that site. This is unlike the situation for an uncorrelated electron, which can  always hop,  regardless of  the opposite spin occupancy of the target site.
Hence the motion of a correlated  electron of either spin is accompanied by the ``backflow'' of a vacancy. It is  therefore clear that the carriers are best viewed  as holes {\em measured from half filling}.  Thus at a filling of electrons $n \equiv N/N_s $ (where N and $N_s$ are the total number of electrons and the number of lattice sites), the carriers are in fact $\delta= |1-n|$, so that near  half filling $\delta \rightarrow 0$, and one sees that the carriers are frozen out. Thus the overall scale of several transport coefficients can be found almost  by inspection; e.g.  the Hall  number must vanish as we approach half filling, as must the  inverse thermopower, defined below in Eq(\ref{thermopower}).    However, it is already clear that the Bloch Boltzmann approach cannot easily capture these ``obvious results''. The latter starts with the band structure derived quasiparticles, and as $n\rightarrow 1$,  has no knowledge of the impending disaster, also known as the Mott insulating state!  One can also view this issue from the point of view of real space versus momentum space definition of holes: the correlated matter clearly requires a real space picture to make physical sense (as opposed to computational ease), whereas the Bloch Boltzmann approach takes a purely momentum space approach to particle and holes. In fact the ``Bloch Boltzmann holes" are vacancies in momentum space measured from a {\em completely filled band}, and have no resemblance to the Mott Hubbard holes. Of course the above diatribe obscures  a crucial point, the Mott Hubbard real space holes view point is almost impossible to compute with, at least using  techniques that exist so far.
  On the other hand,  the momentum space view is seductive because of the ease of computations exploiting  a well oiled machine, namely the perturbative many body framework. Hence it  seems profitable to explore methods and techniques that implement the Mott Hubbard correlations at the outset, and  give qualitatively correct answers. Our formalism, described below, was motivated by  these considerations.  

In this article, there will be little effort at an exhaustive literature survey.
However, it is appropriate to mention that the problems discussed here have been addressed by several authors recently. Mahan's  articles\cite{mahan_review}  address issues in low carrier density  thermoelectric materials, including superlattices.    Dynamical mean field theory, reviewed in\cite{dmft},   has been applied to the problem of thermoelectricity  in\cite{kotliar}. A considerable body of theoretical and experimental work on heavy Fermi systems and relevant models can be found in the work\cite{zlatic}. In particular the review article\cite{zlatic_rmp} summarizes  the work on  the Falicov Kimball model as an application of the dynamical mean field theory.

  Our published papers contain more references  to other approaches taken in literature. I would however, like to mention that at a ``mean field theory'' level, the Mott Hubbard correlations {\em can be built in}, by various slave Boson or slave Fermion approaches, with some success \cite{lee}. In essence, strong correlations force us to deal with Gutzwiller projection Eq(\ref{gutzwiller}) of  the Fermi operators
\begin{equation}
\hat{c}_{j \sigma} = P_G \ c_{j \sigma} \ P_G,
\end{equation}  
and a similar expression for the creation operators. The sandwich of the operators by $P_G$ 
makes sure that the states considered have no double occupancy, $P_G$ annihilates those states.
However,  the operators $\hat{c}_{j \sigma}$ are no longer canonical Fermions, i.e. do not satisfy the usual anticommutation relations.   One finds  that  $\{ \hat{c}_{j \sigma} , \hat{c}^\dagger_{l \sigma} \} \neq \delta_{j,l}$, but rather a non trivial term appears on the right hand side. One way to avoid dealing with the $\hat{c}_{j \sigma}$ operators is to  
  represent the effects of the Gutzwiller projection, using auxiliary (``slave'') Fermi or Bose operators to force the constraint of no double occupancy\cite{lee}. These slave fields consist of canonical Fermions or Bosons, but with an added constraint at each site, and in order to deal with that constraint end up making  Hartree type factorization of resulting expressions. The errors made by these factorizations are hard to quantify, but do give some qualitative understanding of transport in many cases.

Since well controlled calculations are difficult to perform for the experimentally relevant case of 2-dimensions with electrons having spin $\frac{1}{2}$, we are most often forced into numerical computations. The formalism developed here provides some guidance  towards effective computations. We expect that our formalism is  to be  supplemented by a heavy dose of numerics, either exact diagonalization or some other means. 
\subsection{ Plan of the article.}
 In Sec.2,  we motivate the high frequency approach through the example of the Hall constant.  For the triangular lattice sodium cobaltate, this  leads to the  interesting  prediction  of a $T$ linear Hall constant, which has been verified experimentally.  In Sec.3 we   obtain the  finite frequency  thermoelectric  response functions, by using a dynamical version of  Luttinger's gravitational  field as a proxy for the thermal gradients. From this formalism,  novel sum rules for the thermal conductivity  and new thermal and thermoelectric operators emerge.  We obtain  useful formulas for the variables of common  interest such as the Seebeck coefficient and the figure of merit. In Sec.4 we  present the result of applying   these formulas numerically to sodium cobaltate, and benchmark the high frequency approximation by comparing with the exact evaluation of Kubo's formulas. We  show how our formalism gives a quantitatively accurate result for  existing materials. It further 
leads to interesting, and possibly  important predictions for the  Seebeck coefficient of as yet undiscovered materials. In Sec.5  we present a simple diffusion  relaxation model for  coupled charge and heat currents in metals, where the new operators play an explicit role, and their meaning is made physically clear. The  model and some novel response functions relating to an applied AC power source, are likely to be of interest in the context of pulsed laser heating in materials.

\section{Hall constant}
 The basic idea of this approach is well illustrated by the example of the Hall constant for correlated matter $R_H$ defined in Eq(\ref{hall}). Here the initial paper of Shastry, Shraiman and Singh \cite{ssshall} pointed out that the {\em dynamical Hall constant} is better suited for computation in correlated  systems. Consider the simplest framework, the Drude theory of electrons\cite{ashcroft_mermin,ziman}, where we know that 
\begin{eqnarray}
\sigma_{xx}(\omega)&=&  \frac{ \sigma_{xx}(0)}{( 1+ i \omega \tau)}\nonumber \\
\sigma_{xy}(\omega)&=& \frac{ \sigma_{xy}(0)}{( 1+ i \omega \tau)^2} \nonumber \\
B \ R_H \equiv  \rho_{xy}(\omega)& = &   \frac{\sigma_{xy}(\omega)}{ \sigma_{xx}(\omega) \; \sigma_{yy}(\omega)} =  \frac{B}{n q_e c},  \label{hall}
\end{eqnarray}
where $q_e= - |e|$ is the electron charge, $n$ the density of electrons and $\tau$ the relaxation time and $B$ the uniform magnetic field along the z axis. The relaxation time cancels out in computing the Hall resistivity at arbitrary frequencies, and this cancellation gives us a clue. We might as well compute the two conductivities $\sigma_{\alpha \beta}(\omega)$ at {\em high frequencies}, since here the notorious difficulties inherent in computing the DC values of these objects vanish. The Drude theory therefore gives us an important insight, namely that the {\em Hall resisitivity}
is less $\omega$ dependent than the {\em Hall conductivity}. We explore and build on this central idea further in this article, using exact diagonalization, dispersion relations and sum rules.

In order to perform the above suggested  calculation, we need to take the Kubo formulae for the conductivities\footnote{ It is frustrating that despite several ambitious claims in literature, especially from the Mori formulation experts,  there is no practical and  direct way of computing the dynamical {\em resistivity} that bypasses the intermediate stage of computing the dynamical {\em conductivities}\cite{argyres}.},  and take the appropriate ratios to get the dynamical resistivity. Let us  consider the electrical conductivity $\sigma_{\alpha \beta}(\omega)$ of a general Fermionic system  defined  on a lattice. Let us define  an  energy dispersion $\varepsilon_{k}$ obtained by Fourier transforming the hopping matrix element $t(\ven)$ as $ \varepsilon_{k} = - \sum_{\ven} \ \exp{-i \vec{k}. \ven } \; t(\ven) $.  The electrical current operator  is obtained using  the continuity equation as 
\begin{equation}
\vec{\hat{J}} =  i\ q_e \  \sum_{\vec{x},\ven} \ t(\ven) \ \ven \ c^\dagger_{\vec{x}+\ven,\sigma} c_{\vec{x}, \sigma}. \label{current}
\end{equation}
 The current operator $\hat{J}_\alpha$ is dressed by a suitable Peierls\cite{peierls} phase factor in the presence of the uniform magnetic field B along the z axis. In the $t$-$J$ model, the current is sandwiched by the Gutzwiller projector Eq(\ref{gutzwiller}) as $\hat{J} \rightarrow P_G  \hat{J} P_G$, and thereby allows transport only between singly occupied sites.
We can use perturbation theory to linear order in the external electric field to find a 
 general expression for the dynamical  conductivity\cite{mahan, ssshall}: 
\begin{equation}
\sigma_{\alpha \beta}(\omega_c)= \frac{  i}{\hbar N_s v \omega_c }  \left[ \langle \tau^{\alpha \beta} \rangle   +
 \hbar \sum_{n,m} \frac{ p_n- p_m }{\epsilon_n - \epsilon_m + \hbar \omega_c} \langle n|  
\hat{J}_\alpha | m  \rangle \langle  m |  \hat{J}_\beta | n  \rangle \right], \label{kubo}
\end{equation}
where $p_n \propto e^{- \beta \epsilon_n}$ is the probability of the state $n$, and  the ``stress tensor" (sometimes called the ``effective mass tensor'') is defined by
\begin{equation}
\tau^{\alpha \beta}= q_e^2 \sum_{k,\sigma} \frac{d^2 \varepsilon(k)}{d k_\alpha d k_\beta} c_\sigma^\dagger(k) c_\sigma(k),
\end{equation}
where  $v$ is the atomic volume, and $\omega_c = \omega + i 0^+$. The Hall conductivity infact involves the antisymmetric part of this tensor \cite{ssshall}.
 In the case of a $t$-$J$ model   the $\tau$ operators  are also sandwiched by  Gutzwiller projection Eq(\ref{gutzwiller}). In order to compute say  the transport conductivity  $\Re e \ \sigma_{xx}(\omega)$ in the limit $\omega \rightarrow 0$,  we need to sum over terms such as 
$ \sum_{n,m}  p_n \ \delta ( \epsilon_n - \epsilon_m ) \langle n |  \hat{J}_\alpha | m  \rangle \ \langle  m |  \hat{J}_\beta | n  \rangle  $. Such a computation is made very difficult  by the presence of the Dirac delta functions. These energy conserving delta functions lead to a finite limit for $\sigma^{xx}(0)$ in say a disordered metal. The limit is reached only in the thermodynamic limit by a   subtle limiting process, and corresponds to a dissipative resistivity. These delta functions  are very hard to deal with, if we are  given a set of energy levels for a finite system.  It is then  necessary to  broaden the delta functions to a suitable function, say a Lorentzian with an appropriate width determined by the system size and other parameters. In practice this task is quite  formidable 
and only rarely has it been undertaken, thereby motivating the search for alternate routes.

Following the hint contained in the Drude formulae, we can take the high frequency limits 
for the conductivity and thereby obtain the Hall resistivity at high frequencies \begin{equation}
  R^*_H \equiv \; \lim_{\omega \rightarrow \infty} R_H(\omega) \ = \frac{- i N_s v}{B \hbar} \frac{ \langle [\hat{J}_x,\hat{J}_y] \rangle}{\langle \tau^{xx} \rangle ^2}.\label{rhstar}
\end{equation}
In deriving this formula, one is  working in the non dissipative (reactive) regime. That is because the Kubo formulas in Eq(\ref{kubo}) are evaluated away from the $\omega \rightarrow 0$ limit, where the Dirac delta functions come into play.

  The main article of faith is the claim that $\rho_{xy}(\omega)$ at large frequencies is related in a simple way to the transport variable $ \rho_{xy}(0)$. Is this  rationalizable? Further, what is meaning of high frequency, or how ``high'' is ``high enough''? 

With regard to the magnitude of the frequency, the key point is to work with a projected Fermi system rather than a bare one. For example in the case of the Hubbard model versus the $t$-$J$ model, one sees that the energy scale inequality requirement  is 
\begin{eqnarray}
\hbar \omega & \gg & \{ |t|, U \}_{max} \label{hub1} \\ 
\hbar \omega & \gg & \{ |t|, J \}_{max} \label{tJ1}.
\end{eqnarray}
Thus in case of the $t$-$J$ model, one can be in the high frequency limit, and yet have a modest value of $\omega$, in contrast to the Hubbard model since usually $U$ is large, $O( ev's)$. In case of the cobaltates, the energy scale that determines the  high frequency limit is presumably the Hunds rule or crystal field energy, and hence much lower.  Thus the ``high frequency limit'' is expected to be close to the transport values, for models where the high energy scale is projected out to give an effective low energy Hamiltonian with suitably projected operators. 

Subsequent studies show that this simple formula Eq(\ref{rhstar}) is a particularly useful one, we list some of its merits:
\begin{enumerate}
\item It is exact in the limit of simple dynamics, as in the Bloch Boltzmann equation approach.
\item It can be  computed in various ways, e.g. using exact diagonalization\cite{haerter_shastry_hall_2007} and  high T expansions\cite{ssshall,kumar_shastry}. 
\item We have successfully removed the dissipational aspect of the Hall constant here and retained the (lower  Hubbard sub band physics) correlations aspect. This is done  by going to high frequencies, and  using the Gutzwiller projected Fermi operators in defining the currents.  
\item It is valid for the entire range of hopping processes, from  hopping  type  incoherent transport at high T, to coherent Fermi liquid  type transport at low T in a  band system.
\end{enumerate}
We emphasize that this provides a very good description of the  $t$-$J$ model, where this asymptotic formula  requires $\omega$ to be larger than J, but should not be expected to be   particularly useful for Hubbard model. In the Hubbard model \cite{imada},  the transport limit and the high frequency limit are on opposite sides of a crucial energy scale $U$.
More explicitly, a  large $\omega \gg U$ is implicit in this limit, and  therefore deals with weakly renormalized particles. We  expect it to differ from the transport limit $\omega \rightarrow 0$ significantly in qualitative terms, such as the signs of carriers and the Hall number.

\begin{center}
\begin{figure}[h]
\includegraphics[width=9cm]{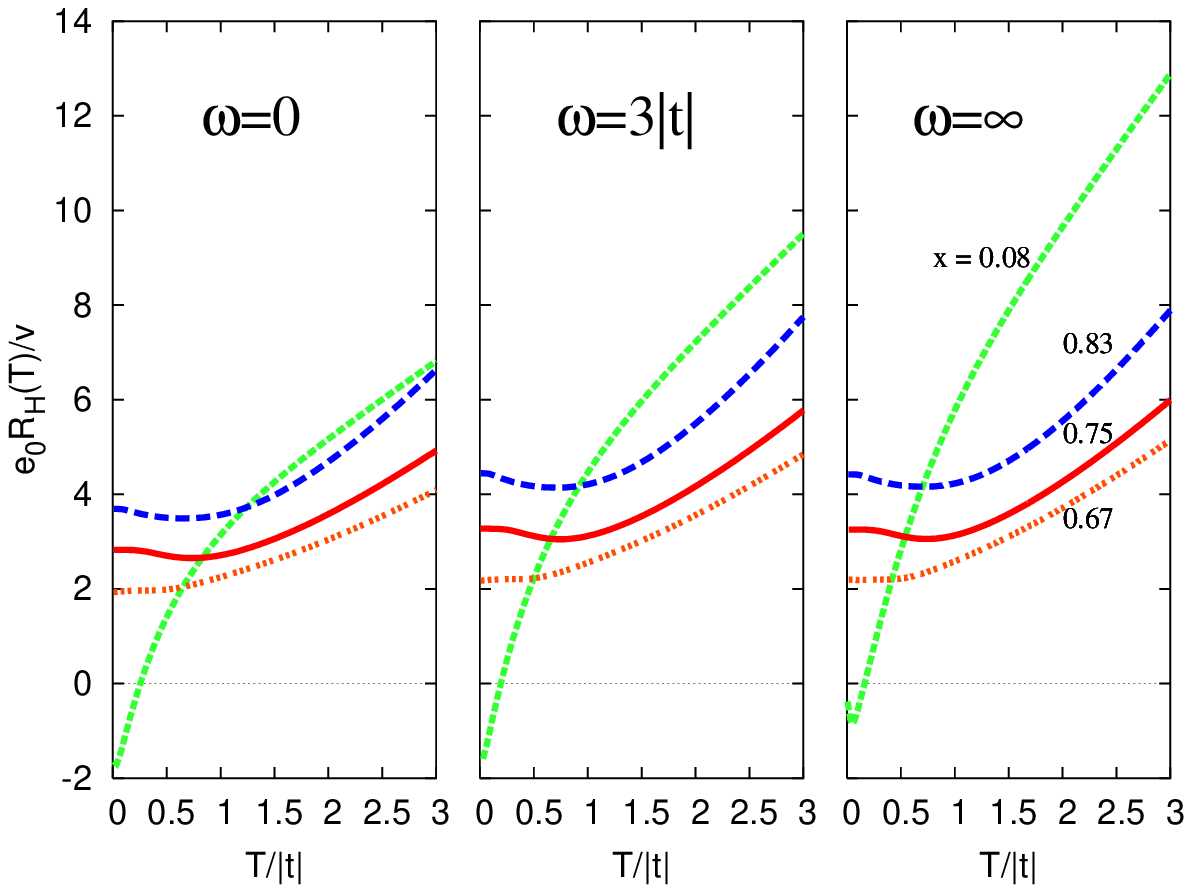} \\
\includegraphics[width=6cm]{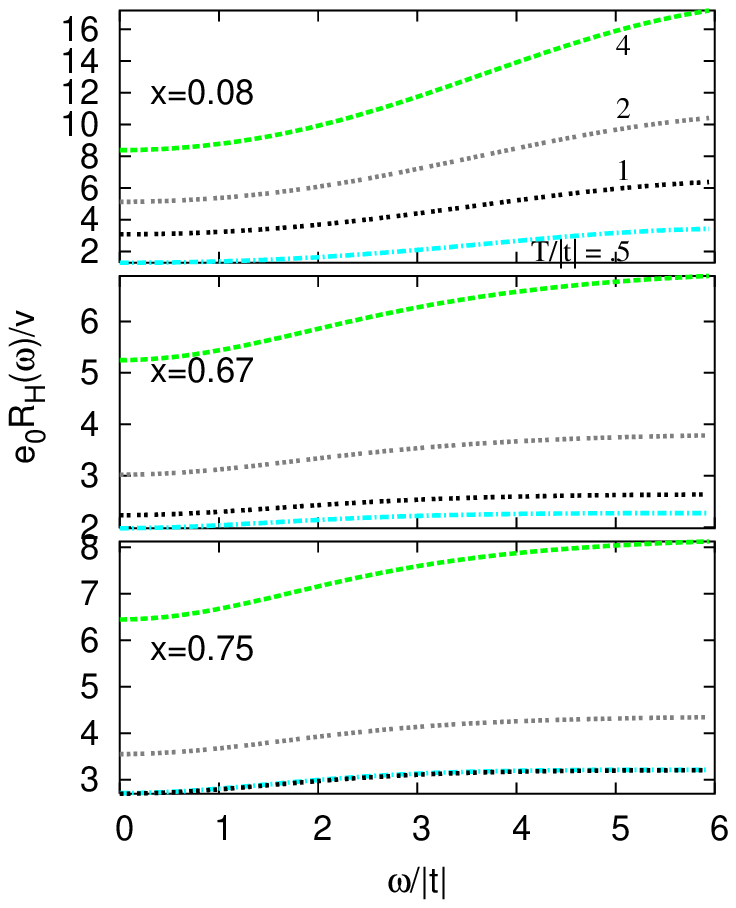}
  \caption{ Frequency-dependence of the Hall coefficient on the triangular lattice  from computation on small clusters of the $t$-$J$ model Ref\cite{haerter_shastry_hall_2007,haerter_peterson_shastry_curie_weiss} for electron doping  $x$. The values of doping $x$ are indicated in the figures. In the upper figure, the linear $T$ dependence is striking in all cases. The bottom figure displays the  frequency dependence for various values of $x$ and $T$.  It is seen from these curves that frequency dependence is modest except for the case of very low doping.}
\label{frequency_dependence_rh}
\end{figure}
\end{center}

It is worthwhile recording a  dispersion relation for the Hall constant at this point. Since $ R_H(\omega)$ is analytic in the upper half of the complex $\omega$ plane, and has a finite limit at infinite $\omega$, we may write
\begin{equation} 
R_H(\omega) = R_H^* - \int_{-\infty}^{\infty} \frac{ d \nu}{\pi}\; \frac{ \Im m R_H(\nu)}{\omega - \nu + i 0^+},  
\end{equation}
therefore in the DC limit we get:
\begin{equation}
 {\Re e}R_H(0)= R_H^*+ \frac{2}{\pi}\int_0^\infty\frac{{\Im m} R_H(\nu)}{\nu}\;d\nu \;. \label{hall_dispersion}
\end{equation}
This equation  quantifies the difference between the experimentally measured DC-Hall coefficient and the theoretically more accessible infinite frequency limit. The second term  is  an independently measurable object, and initial measurements of this are now available in Ref.\cite{drew}. It would be very useful to make a systematic study of this promising  dispersion relation, both theoretically and experimentally.
For the case of the square lattice systems, the theoretical estimates of the difference are  indicated in Fig.(\ref{frequency_dependence_rh}) for a couple of densities. We plan to return to this  rich topic   in future studies. 
As an illustration of the above formalism, we note that  recent work on triangular lattice system $Na_xCoO_2$ provides a good example. Theoretically, the ``exotic'' prediction, namely that  the Hall constant grows  linearly with temperature T on a triangular lattice, was first recognized in 1993 Ref\cite{ssshall}.  This behaviour arises for $T \geq T_{Fermi}$. On the other hand,  for low temperatures, it is expected to be less sensitive to $T$, as in a Fermi liquid. This prediction arises in a simple way  from Eq(\ref{rhstar}) treated within the high temperature expansion\cite{ssshall,kumar_shastry}. The numerator is dominated by the shortest closed loops of electron hopping that encircle a flux, and these are of course triangles for the triangular lattice. This leads at high $T$ (or small inverse temperature $\beta$) to the numerator $\propto \beta$ whereas the denominator is always $\propto \beta^2$, and hence a $T$ linear Hall constant with a well defined coefficient
\begin{equation}
R_H^*=- \frac{v}{4 |q_e|} \frac{k_B T }{t} \frac{1+\delta }{\delta( 1- \delta)} + c_1+  \frac{c_2}{T}+\cdots .\label{tlinearhall}
\end{equation}
This result (with suitable constants $c_1, c_2$) Ref\cite{kumar_shastry}, is for the experimentally relevant case of electron filling so that $\delta= \frac{N}{N_s}-1$, and has a suitable counterpart for the case of hole filling. It is remarkable in two distinct ways. Firstly, it shows that the sign of the Hall constant is not universal, as one might naively  expect from the Sommerfeld Drude theory formula $R_H=1/n q_e c$.   Rather it depends upon the {\em sign of the hopping} as well. This aspect was recognized in the  important work of
Holstein\cite{holstein}, within the context of hopping conduction in doped semiconductors.
The other remarkable feature is that the Hall resistivity increases linearly with $T$, a result first found in \cite{ssshall, kumar_shastry}\footnote{ Using   a semiclassical theory of transport, Holstein estimated the Hall conductivity and Hall angle $\sigma_{xy}/\sigma_{xx}$, rather than the Hall resistivity as in Eq(\ref{tlinearhall}). The neat prediction\cite{ssshall, kumar_shastry} of a $T$ linear behaviour Eq(\ref{tlinearhall})  emerges only for the Hall resistivity, where many  factors cancel out.} . The final answer is therefore highly non universal, and depends upon material parameters such as the magnitude and sign of the hopping, and also the nature of the doping (holes versus electron). We  reiterate   that this asymptotic  behaviour is obtained provided $k_B T \geq |t|$ and as such is experimentally observable   only for  narrow band systems. In general, from Eq(\ref{rhstar}) one expects a $T$ independent Hall constant for  $T$ sufficiently below a (usually large) characteristic Fermi temperature, as in most metallic systems. 
\begin{center}
\begin{figure}[h]
\includegraphics[width=9cm]{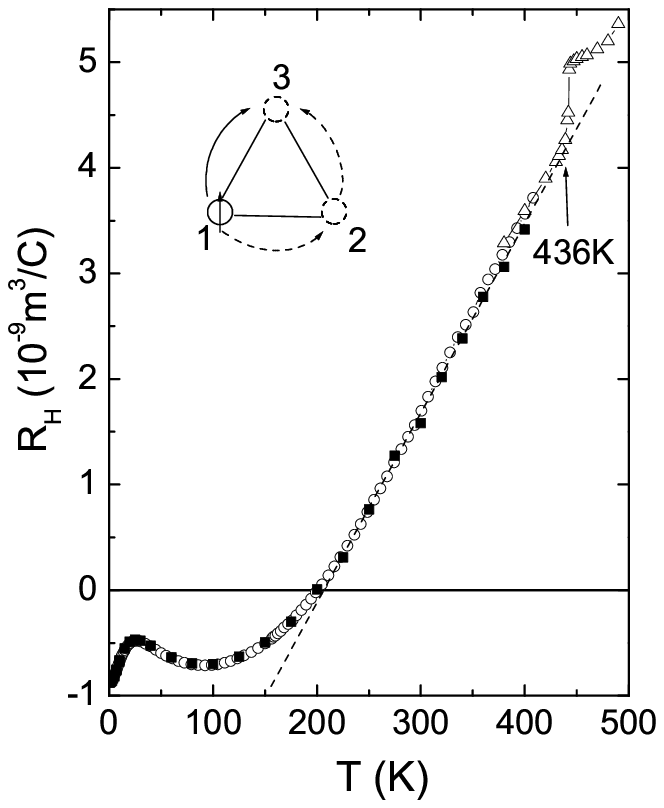}
  \caption{ Experimental temperature dependence Ref\cite{wang_ong} of the Hall coefficient of sodium cobaltate  $Na_{.68}CoO_2$ over a broad range of temperature. The sample is in the  so called Curie Weiss metallic phase. The inset stresses the crucial role of the triangular closed loops in giving rise to the surprising behaviour.}
\label{tlinear_rh}
\end{figure}
\end{center}
Interestingly enough, the case of $Na_xCoO_2$ with $x\sim 0.68$, i.e. the so called Curie Weiss metallic  phase, seems to fulfill these conditions of narrow bandwidth.  As Fig.(\ref{tlinear_rh}) shows, the experiments show a large and clear-cut  region of  linear T dependence\cite{wang_ong}, thereby fulfilling the basic theoretical prediction of Eq(\ref{tlinearhall}). Recent work\cite{haerter_peterson_shastry_curie_weiss} attempts to reconcile many experimental results in this phase, including  the Hall constant coefficient of $T$, with the theoretical predictions.  Many experiments such as the photoemission quasiparticle velocities, the magnetic susceptibility and specific heat are understandable with $|t|/k_B \sim 100^0$K (i.e. a bare band width $ 9 |t| \sim 10^{-2}$ eV ).  At $x=.68$, the Hall  slope requires a smaller value  $|t|/k_B \sim 25^0$K, but nearby compositions seem to have a smaller slope translating to a larger value of $|t|/k_B$ that is more in line with the other data.
All these numbers are, in turn, much smaller than LDA estimates of the bandwidth $0.2$ eV \cite{singh} by an order of magnitude, and pose an interesting  problem to the community. In this article, our interest in the Hall constant of the cobaltates is mostly motivational and hence  tangential; we will leave this topic for further work. In the case of the cuprates, the  work in Refs \cite{ssshall,haerter_shastry_hall_2007}  shows that  $R_H^*$ provides a useful first principles estimate for the physical (DC) transport Hall constant $R_H(0)$ for correlated systems. Our task in these notes is to carry this message to the computation of the thermal response functions,  and so we terminate  our discussion of the interesting problem of the Hall constant.

\begin{figure}[h]
\includegraphics[width=9cm, angle=-90]{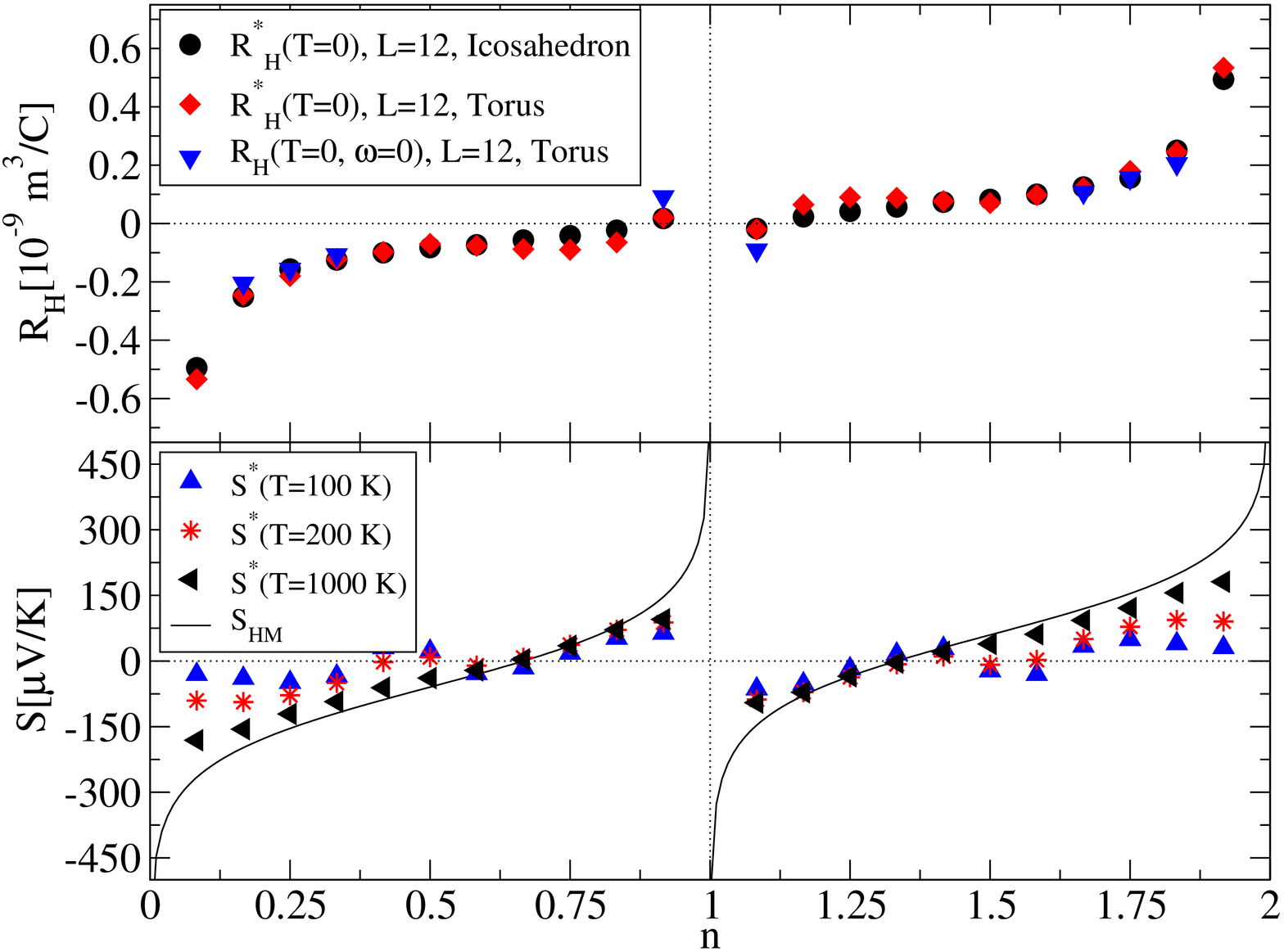}
\caption{In a Mott Hubbard system both the Hall constant and the Seebeck coefficient have {\em three} zero  crossings as the band is populated from $0 \leq n \leq 2$. The divergence at half filling is weaker in the Seebeck coefficient than in the Hall constant, as shown in this
example from the $t$-$J$ model on the triangular lattice Ref\cite{haerter_shastry_hall_2007,haerter_peterson_shastry_curie_weiss}. The three zero crossings are in contrast to a single zero crossing of an uncorrelated band.  The distinction   is   understood as a consequence of the Mott Insulating state at half filling\cite{ssshall,shastry_sumrule}. This insulating state determines
the physics  of the carriers in its proximity, and these are argued here to be far from the Bloch Boltzmann holes of standard transport theory.  The location of the zero crossings is determined by  details such as the lattice structure.}
   \label{hall_thermopower}
\end{figure}

\section{Thermoelectric  Response}
We next address the main topic of this  article, namely the thermal response functions. In light of the previous discussion of the Hall constant,  we searched  for the analog of $R_H^*$. Therefore we needed  the finite  (high) frequency limits of thermal response functions. To the author's surprise, these limiting functions were unavailable in literature, therefore leading to the basic calculation in\cite{shastry_sumrule}. We begin with a quick review of the standard transport theory given in many texts\cite{onsager,luttinger,ashcroft_mermin,ziman,mahan}. We write down  the set of linear response equations following Onsager\cite{onsager}
\begin{eqnarray}
\frac{1}{\Omega} \langle \hat{J}_x \rangle &=& L_{11} E_x + L_{12}  (- \nabla_x T / T) \label{onsager1}  \\
\frac{1}{\Omega}  \langle \hat{J}^Q_x \rangle &=& L_{21} E_x +L_{22}  (- \nabla_x T /T), \label{onsager2}
\end{eqnarray}
where $(- \nabla_x T/T) $ is regarded as the {\em external driving thermal force}\cite{kelvin,onsager,ashcroft_mermin}. The operator  $\hat{J}_x$ is the total charge current operator, and  has been defined earlier in Eq(\ref{current}). Further  $\hat{J}^Q_x$ is  the     heat current operator defined as $\hat{J}^Q_x= \lim_{q_x \rightarrow 0} \frac{1}{ q_x} \left[ K, K(q_x) \right]$, where $K(q_x)$ is the Fourier component of the Grand canonical Hamiltonian density Eq(\ref{ktotal}), and $\lim_{q_x \rightarrow 0} K(q_x) =K$. These variables are  elaborated upon  below in Eq(\ref{heat_current},\ref{heat_current_2}) and ${\Omega}= v N_s $ is the total volume of the system. The parameter $L_{11}$ is related to the DC conductivity $\sigma(0)=L_{11} $\footnote{Our definition includes the volume factor and this makes $L_{11}$ identical to the  (intensive) conductivity.},  the parameter $L_{12}$   is related\footnote{ Sometimes  in literature \cite{ashcroft_mermin,ziman,mahan}, S is denoted by Q.} to  the Seebeck coefficient 
\begin{equation}
S=\frac{L_{12}}{T L_{11}},\label{thermopower}
\end{equation}
  also $L_{21}$ is related to  the Peltier coefficient 
\begin{equation}
\Pi=L_{21}/L_{11} = T S   \label{peltier},
\end{equation}
the final equality in Eq(\ref{peltier})  
relating the Peltier and Thomson effects is    the celebrated  reciprocity due to  Thomson (Kelvin)\cite{kelvin} and Onsager\cite{onsager}. It  is most compactly written as 
 $L_{12}=L_{21}$.
 The Onsager constant  $L_{22}$ is related to  the (nominal) thermal conductivity $\kappa=\frac{1}{T} L_{22} $ for  problems with immobile degrees of freedom (spins, ions, etc). For metallic systems, however,  the observed  thermal conductivity $\kappa_{zc}$ requires a  small correction (see Eq(\ref{thermal_zero_current})).  The usually observed  thermal conductivity\cite{ziman,mahan,peierls} uses  the {\em zero electrical current condition} $\langle \hat{J}_x \rangle =0$, thereby inducing an electric field.    The  generated electric field is related by Eq(\ref{onsager1}) to the applied thermal force, and using it in  Eq(\ref{onsager2}) we find the zero current thermal conductivity\cite{mahan,ziman} 
\begin{equation}
\kappa_{zc}= \frac{1}{T  L_{11}}( L_{22} L_{11} - L_{12} L_{21} ). \label{thermal_zero_current}
\end{equation}
 These are equations in the static limit, and correspond to the most simple non equilibrium states with a steady current flow.
 
\subsection{Luttinger's gravitational field analogy}
In order to generalize the above transport theory  to finite frequencies, we need to borrow  a  beautiful idea from Luttinger Ref\cite{luttinger}.  In order to derive the  Kubo formulae Ref\cite{kubo}, he introduces the   mechanical equivalent of the thermal gradient, and we shall use it extensively.
The fictitious  mechanical field  $\psi(\vec{x},t)$ is  similar to a gravitational field,   coupling to the  effective ``mass density'' $m_{eff}(\vec{x})= \frac{1}{c^2} K(\vec{x})$ via 
\begin{equation}
K_{tot} = K +   \sum_x K(\vec{x}) \psi(\vec{x},t). \label{ktotal}
\end{equation}  
Here $K= \sum_x K(\vec{x})$, and  $K(\vec{x}) = H(\vec{x}) - \mu n(\vec{x}) $  is the {\em Grand canonical} Hamiltonian\footnote{The need of introducing the Grand canonical Hamiltonian K lies in the construction of the heat current operator $\hat{J}^Q_x$, where the particle current must be subtracted from the energy current Eq(\ref{heat_current}).},  $H(\vec{x}), n(\vec{x}) , \mu$ are the local  canonical ensemble Hamiltonian, number density and chemical potential.
Below, we will expand $K(\vec{x})= \frac{1}{\Omega} \sum \exp{ - i \vec{q}.\vec{x}}\ K(\vec{q})$, with a similar expansion for the charge and other densities and currents.
 We can compute the standard linear response to a space time dependent $\psi(\vec{x},t)$, and with the help of the ideas initiated by Luttinger, deduce the dynamical thermal  response functions required in Eq(\ref{lresponse}).

 Firstly let us note that the local temperature $\delta T(\vec{x},t)$ can be defined in the long wavelength almost static limit through small departures from equilibrium.  The local energy fluctuation can be written as  $\langle  K(\vec{x},t) \rangle
= \langle K \rangle_0 + C(T) \ \delta T(\vec{x},t)$, with $C(T)$ as the specific heat at the equilibrium temperature $T$(at constant volume and $\mu$), provided $\delta T(\vec{X},t) \ll T$.  Hence we can invert to define the local temperature through
\begin{equation} \delta T(\vec{x},t) = \frac{\delta \langle K(\vec{x},t)\rangle}{C(T)}. \label{deltat} 
\end{equation}
The connection of $\psi(\vec{x},t)$  with local temperature $\delta T(\vec{x},t)$ emerges from a  study of the generalized phenomenological equations proposed by Luttinger\cite{luttinger}. He specializes to long wavelength $\vec{q} \rightarrow 0$ and static  $\omega \rightarrow 0$ limits where equilibrium is rigorously definable;  we will  extend this notion  to arbitrary variations. The phenomenological relations  are generalizations of the Onsager formulation\cite{onsager} as in Eq(\ref{onsager1}) and Eq(\ref{onsager2}), and correspond to adding terms proportional to the gradient of the mechanical term $\propto \psi$ in Eq(\ref{ktotal}). Luttinger  writes
\begin{eqnarray}
\frac{1}{\Omega} \langle \hat{J}_x \rangle &=& L_{11} E_x + L_{12}  (- \nabla_x T / T) + \hat{L}_{12}  (- \nabla_x \psi(\vec{x},t))\label{luttinger1}  \\
\frac{1}{\Omega}\langle \hat{J}^Q_x \rangle &=& L_{21} E_x + L_{22}  (- \nabla_x T /T) + \hat{L}_{22}  (- \nabla_x \psi(\vec{x},t)), \label{luttinger2}
\end{eqnarray}
where the two new response functions $\hat{L}_{12}, \hat{L}_{22}$ are  functions of space and time which can be readily computed from a linear response theory treatment  of the mechanical perturbation in Eq(\ref{ktotal}).
We will treat $\psi$ as a small perturbation and work to linear order here.
 Addition of the $\psi$ term in these equations allows us to take a different perspective\footnote{ Note that experiments usually employ {\em open boundary conditions}, and the temperature gradient is externally applied. The usual argument made is that the periodic boundary case and the open boundary case are equivalent, provided we take the wave vector $\vec{q}\rightarrow 0$ or the thermodynamic volume $\Omega \rightarrow \infty$ limits respectively, while keeping the frequency $\omega$ finite and small. This gives a prescription for the DC limit in both cases, namely to take the DC limit at the end of the volume (or wave vector) limits Eqs(\ref{onsager1},\ref{onsager2}).} In Eqs(\ref{luttinger1},\ref{luttinger2})  {\em we  can  view the driving term as} $\psi$, with  the temperature fluctuation arising as a consequence of this driving, at least for long wavelengths and slow variations\footnote{ This is where Luttinger uses the tactical analogy with the Einstein relation for the relationship between  self diffusion and conductivity. In the phenomenological equation $\langle \hat{J}_x \rangle  = \sigma E_x + D (-\nabla_x)\langle \rho \rangle $, the driving term is $E_x$. In Eq(\ref{luttinger1}) (neglecting the $L_{11}$ term for a moment), the $\psi$ term is analogous to the $E_x$ in the diffusion problem, and the induced temperature variation is similar to the induced charge fluctuation.
For completeness,  we summarize Luttinger's argument for this case.   For small wave vectors and  slow variation of the electric field  $E_x =  - \nabla \phi(x) = E_0 \exp{- i (q_x x + \omega t)}$.  
Upon using the continuity equation $\langle \rho_q \rangle = - \frac{q_x}{\omega} \; \langle \hat{J}_x \rangle$ we see
 that $\langle \hat{J}_x \rangle = \sigma E_0 \frac{\omega}{\omega +  i D q^2_x}$.     Similarly the charge fluctuation $\langle \rho_q \rangle = \sigma \phi_q \frac{- i q^2_x}{\omega +  i D q^2_x} $, where $\phi_q = - i E_0/q_x$. Luttinger's argument  is that in the fast or transport  limit $\omega \rightarrow 0, q_x \rightarrow 0$ so that the diffusion term can be dropped. However, in the slow limit, the relations derived above show that $\frac{\sigma}{D}= - \langle \rho_q \rangle/\langle \phi_q \rangle$. The right hand side of this is easily computed from {\em thermodynamics},  whereby the Einstein relation $\frac{\sigma}{D}= e^2/(\partial \mu/\partial n)_T$ follows.
}.

In these Eqs(\ref{luttinger1},\ref{luttinger2}), the idea is to determine the difficult unknowns $L_{12},L_{22}$ in terms of the easier objects $\hat{L}_{12}, \hat{L}_{22}$. Let us consider one particular example for simplicity, the others follow similarly. Since the theory is linear in the external perturbation, it suffices to consider a single frequency and wave vector mode. Therefore,
 let us focus on Eq(\ref{luttinger1}), and  introduce a single Fourier component $\psi(\vec{x},t)= \psi_q \exp
\{ -i ( q_x x +  \omega t + i 0^+ t )\} $, (adiabatic switching implied) and the electric potential $\phi(\vec{x},t)= \phi_q \exp \{ -i ( q_x x +  \omega t + i 0^+ t )\}$. We  thus write
\begin{equation}
\frac{1}{\Omega} \delta \hat{J}_x =  L_{11}(q_x,\omega) ( i q_x) \phi_q + ( i q_x) \left[ L_{12}(q_x,\omega) \frac{\delta T_q}{T} + \hat{L}_{12}( q_x,\omega)\; \psi_q \right], \label{psi_1}
\end{equation}  
where $ \langle \hat{J}_x(\vec{x}) \rangle = \frac{1}{\Omega} \delta J_x \exp{ - i( q_x x +  \omega t)} $, so that $\delta{J}_x $   is the amplitude of the response, and we have written the arguments of the Onsager-Luttinger functions $L_{ij},\hat{L}_{ij}$ explicitly. 

To be  explicit, we define two extreme limits of $\vec{q} \ \& \ \omega$ that arise here\cite{luttinger}, one is the so called rapid or transport limit, and the other is the slow or the thermodynamic limit. In the  rapid or transport limit, we first let $q_x \rightarrow 0$ and then let  $\omega$ vanish.
In the slow  limit, we set $\omega \rightarrow 0$ first and then take the limit
$q_x \rightarrow 0$.

 In the transport limit, we  have a spatially uniform  field, and hence we can show that $\delta T_q \rightarrow 0$. This is most easily seen by inspecting the continuity equation for heat density and current in the absence of an external heating source:
$\omega \langle K_q \rangle + q_x \langle  \hat{J}^Q_x \rangle =0$. 
This  can be written, using Eq(\ref{deltat}) as $\delta T_q= \frac{ - q_x}{ C(T)\ \omega} \langle \hat{J}^Q_x \rangle$. Thus dropping the $\delta T_q$ term, we find 
\begin{equation}
\frac{1}{\Omega} \delta \hat{J}_x =  L_{11}(0,\omega) \lim_{q_x \rightarrow 0}( i q_x)  \phi_q + \hat{L}_{12}(0,\omega) \lim_{q_x \rightarrow 0} (i q_x) \psi_q.
\end{equation}
The object $ \lim_{q_x \rightarrow 0}( i q_x)  \phi_q \rightarrow E_x $ and likewise for the gravitational term, and hence this equation is essentiall the same as Eq(\ref{luttinger1}) above.

On the other hand, in taking the slow  limit, with $\omega \rightarrow 0$, the system is  subject to a time independent but a spatially    varying gravitational  potential as well as a temperature gradient; this is now an equilibrium problem without a net current. Thus $\langle \hat{J}_x(q_x) \rangle =0$, leading to 
\begin{equation}
0= L_{12}(q,0) \frac{\delta T_q}{T} + \hat{L}_{12}(q,0) \psi_q. \label{zeroj}
\end{equation}
In this equilibrium situation, we can compute the connection between $ \frac{\delta T_q}{T}$ and $\psi_q$
readily. Using  lowest order thermodynamic perturbation theory Ref\cite{mahan,fetter}
we compute the change in energy induced by a small perturbation $\psi_q$ 
\begin{equation}
\frac{\delta \langle K(\vec{q}) \rangle}{ \psi_q} =   - \sum \frac{p_n-p_m}{\varepsilon_m-\varepsilon_n } |\langle n| K(\vec{q}) | m \rangle|^2 + O(\psi), \label{spht1} 
\end{equation}
with $p_n= \frac{1}{{\cal Z}} \exp( - \beta \varepsilon_n)$ the probability of the state $n$. In the limit $\vec{q} \rightarrow 0$,
  $K(\vec{q})$ tends to the Hamiltonian, and hence cannot mix states  of different energy, hence we write $\lim_{\varepsilon_m \rightarrow \varepsilon_n} \frac{p_n-p_m}{\varepsilon_m-\varepsilon_n } \rightarrow \beta p_n$, whereby
\begin{eqnarray}
\lim_{\vec{q}\rightarrow 0} \frac{\delta \langle K(\vec{q}) \rangle}{ \psi_{q}} & \rightarrow& - \beta \left[  \langle K^2 \rangle - \langle K \rangle^2 \right] \label{spht2}\\ 
&=& - T C(T) \label{spht3} 
\end{eqnarray}
This calculation is parallel  to that in literature\cite{compressibility_sum_rule} for the electron liquid, where the dielectric function is related to the the compressibility in the  limit of $\vec{q} \rightarrow 0 \; \omega \rightarrow 0$. Comparing the final Eq(\ref{spht3}), with the standard thermodynamic definition of $C(T)$, we see that
\begin{equation}
\lim_{q \rightarrow 0} \frac{\delta \langle K(\vec{q}) \rangle}{ \psi_{q}} = - T \frac{d}{d T} \langle K \rangle, \label{spht4} 
\end{equation} 
whereby
\begin{equation}
\lim_{\vec{q} \rightarrow 0}  \psi_{q} = - \lim_{\vec{q} \rightarrow 0} \frac{ \delta T_q}{T}. \label{spht5}   
\end{equation}
Comparing Eq(\ref{spht5}) and Eq(\ref{zeroj}), we see that
\begin{equation}
\lim_{q \rightarrow 0}\left[ L_{12}(q,0)- \hat{L}_{12}(q,0) \right] = 0\label{luttinger_identity}.
\end{equation} 
From this relation, Luttinger concludes that $L_{12}$ in the DC limit can be computed from $\hat{L}_{12}$. Thus the problem of computing thermal response is reduced to computing mechanical response to the field $\psi(\vec{x},t)$, and essentially treating\footnote{The alert reader would have noted that this assignment has an opposite sign from Eq(\ref{spht5}). The explanation of this slight ``booby trap'' is that in Eq(\ref{spht5}), the gravitational field and the thermal gradient 
are simultaneously present in order to  cancel the current. Their relative sign is therefore negative. In making the suggested replacement, the gravitational field is used {\em as a proxy}  for  the temperature gradient, and hence the relative sign is reversed from the earlier context. }  the   $\lim_{\vec{q} \rightarrow 0}  \psi_{q} =  \lim_{\vec{q} \rightarrow 0} \frac{ \delta T_q}{T}$.

 This is undoubtedly huge   progress. However,  as far as I can make out, this fine proof of Luttinger   makes another    implicit assumption, namely  that 
\begin{equation}
\lim_{\omega \rightarrow 0}\left[ L_{12}(0,\omega)- \hat{L}_{12}(0,\omega) \right] = 0,
\end{equation}
somehow follows from Eq(\ref{luttinger_identity}). This is assumed so  despite the fundamental difference in the two limits, namely the slow (thermodynamic) and fast (transport) limits. The belief thus  seems to be that the two functions $L_{ij} $ and $\hat{L}_{ij}$ must be identical in the fast limit, if they are so in the slow limit.

In this work we need to  define finite $q,\omega$ thermal response functions. Towards this end, 
 we will in fact {\em extend the above   to all} $q,\omega$, and simply assume that 
\begin{equation}
L_{ij}(q,\omega) = \hat{L}_{ij}(q,\omega).
\end{equation}
The RHS is computable within perturbation theory, and the LHS, although defined rigorously only in the regime of small $q,\omega$ by hydrodynamic type reasoning, is extended to all $q,\omega$ by this relation. This idea  of extending the thermal functions seems reasonable, since the resulting functions agree with hydro-thermodynamics for small $q,\omega$, and are guaranteed to satisfy general properties such as causality and Onsager reciprocity. With this, we can define all thermal response functions at all $q,\omega$, and in the sequel we will work within  this generalized Luttinger  viewpoint.

\subsection{Finite $\omega$ thermal response functions} 

With this preparation, we  return to exploring the thermal response Eq(\ref{lresponse}) at finite frequencies. The timing of our quest  seems fortuitous, since  there is growing experimental interest in the transport of energy and heat pulses, requiring  knowledge of these variables, and of the approach to equilibrium.

We first need to define the heat current $\hat{J}^Q_x$. Towards this end, we take the time derivative of the first law of thermodynamics for fixed volume $ T \frac{ dQ}{d t}= \frac{ d E}{dt} -\mu \frac{d n}{d t}$. Imagining a small volume with the flow of energy and heat as well as density, and applying this law  locally,  it is reasonable to identify the heat current as the energy current minus the particle current (times $\mu$).  Thus the heat current can  be decomposed as the difference of two terms:
\begin{equation}
\hat{J}^Q_x = \hat{J}^E_x - \frac{\mu}{q_e} \hat{J}_x, \label{heat_current_decomposition}
\end{equation} 
where $\hat{J}^E_x$ is the {\em energy current} and $\hat{J}_x$ the charge current. In a quantum mechanical system, the heat current operator is easiest computed from the  commutator
of the energy density operator with total energy as follows (setting $\hbar =1$): 
\begin{equation}
\hat{J}^Q_x= \lim_{q_x \rightarrow 0} \frac{1}{ q_x} \left[ K, K(q_x) \right]. \label{heat_current}
\end{equation}
This construction is similar to the more familiar one  for  the charge current $\hat{J}_x = \lim_{q_x \rightarrow 0} \frac{1}{ q_x} \left[ K, \rho(q_x) \right]$.
By inspection, a local heat current operator can also be written down provided the interactions are local, so that we can take Fourier components in a periodic box\footnote{We imagine doing this calculation on a lattice, therefore the Fourier transforms are written as sums over sites, with a factor of the atomic volume $v$ inserted for keeping track of dimensions.} and write
\begin{equation}
\hat{J}^Q_x(\vec{q})= v \sum_x \hat{J}^Q_x(\vec{x}) \exp( i \vec{q}.\vec{x}), \;\;\mbox{and}\;\; \hat{J}_x(\vec{q})= v \sum_x \hat{J}_x(\vec{x}) \exp( i \vec{q}.\vec{x}). \label{heat_current_2}
\end{equation}
Therefore, $ \hat{J}^Q_x =\hat{J}^Q_x(\vec{0})$ and $ \hat{J}_x =\hat{J}_x(\vec{0}).$
For different models, the heat current is easy to compute using the above prescription, and many standard models are treated in \cite{shastry_sumrule}.

 Let us impose fields that vary as $\psi(\vec{x},t)= \psi_q \exp \{- i (q_x x + \omega t + i 0^+ t)\}$, and similarly for the electric field with the electric potential $\phi(\vec{x},t)= \phi_q \exp \{-i ( q_x x +  \omega t + i 0^+ t )\}$. Using 
the notation $ \langle \hat{J}_x(q_x) \rangle=  \delta {J}_x $ and
$   \langle \hat{J}^Q_x(q_x) \rangle =  \delta {J}^Q_x  $, we find from 
Eq(\ref{luttinger1},\ref{luttinger2}) 
\begin{eqnarray}
\frac{1}{\Omega} \delta {J}_x &=& L_{11}(q_x,\omega) (i q_x  \phi_q) +  L_{12}(q_x, \omega) (i q_x   \psi_q)  \\
\frac{1}{\Omega} \delta {J}^Q_x &=& L_{21}(q_x, \omega) (i q_x  \phi_q) + L_{22}(q_x, \omega) (i q_x   \psi_q).  \label{lresponse}
\end{eqnarray}
These responses are to be computed for a  Hamiltonian perturbed by a single Fourier component as
\begin{equation}
K_{tot}= K + \left[  \rho(-q_x) \phi_q + K(-q_x) \psi_q \right] \exp{ (-i \omega t +0^+ t)}, \label{hpert1}
\end{equation}
where $\rho(\vec{q})$   is the charge  density fluctuation operator at wave vector $\vec{q}$.

We can reduce the  calculations of all $L_{ij}$  to essentially a single one, with the help of some notation. Keeping $q_x$ small but non zero, we define currents, densities and forces in a matrix notation as follows:
\begin{eqnarray}
  & \mbox{i=1} & \mbox{i=2 } \nonumber \\
 & \mbox{Charge} & \mbox{ Energy} \nonumber \\
 {\cal{I}}_i & \hat{J}_x(q_x) & \hat{J}^Q_x(q_x) \nonumber \\
 {\cal{U}}_i & \rho(-q_x) & K(-q_x) \nonumber \\
 {\cal{Y}}_i & E^x_q= i q_x \phi_q & i q_x \psi_q. \label{notation_1} 
\end{eqnarray}
The perturbed Hamiltonian Eq(\ref{hpert1}) can then be written as
\begin{equation}
K_{tot}= K + \sum_j Q_j e^{ -i \omega_c t}, \;\;\;\mbox{where}\;\;\;
Q_j= \frac{1}{ i q_x} {\cal{U}}_j {\cal{Y}}_j.
 \label{hpert2}
\end{equation}
We denote $\omega_c= \omega + i 0^+$ above and elsewhere. From standard linear response theory\cite{luttinger} applied to Eq(\ref{hpert2}), we readily extract the induced current response
\begin{equation} 
\langle {\cal I}_i \rangle = - \sum_j \chi_{{ \cal {I}}_i , Q_j}(\omega_c), \label{response1} 
\end{equation}
where the susceptibility for any two operators $\chi_{A,B}(\omega_c)$ can be expressed as (with $A_{nm} \equiv \langle n|A| m \rangle$)
\begin{eqnarray}
\chi_{A,B}(\omega_c) & = & i \int_0^\infty \ dt \ e^{ i \omega t -0^+ t} \langle \left[ A(t),B(0) \right] \rangle \nonumber \\
& = &\sum_{n,m} \frac{p_m-p_n}{\varepsilon_n-\varepsilon_m+ \omega_c} A_{nm} B_{mn} \nonumber \\
&=& - \frac{1}{\omega_c} \left[ \langle [A,B] \rangle + \sum_{n,m} \frac{p_m-p_n}{\varepsilon_n-\varepsilon_m+ \omega_c} A_{nm} ([B,K])_{mn}
\right]. \label{chidef}
\end{eqnarray}
The last line of Eq(\ref{chidef}) follows from  integration by parts of the first line, and the average $\langle \rangle$ is carried out over the ensemble where the external fields are dropped.

From Eq(\ref{response1}), using the notation in Eqs(\ref{notation_1},\ref{chidef}), the generalized  Onsager coefficients
\begin{equation} L_{ij}(q_x,\omega) = \frac{1}{\Omega} \lim_{{ \cal{Y}}_j \rightarrow 0} \langle { \cal {I}}_i \rangle / { \cal{Y}}_j  \label{definitionL}
\end{equation}
are written down immediately  
\begin{equation}
L_{ij}(q_x,\omega)  =  \frac{ 1}{ i \Omega \omega_c }  \left[ \langle [{ \cal {I}}_i,{ \cal {U}}_j] \rangle \frac{ 1}{  q_x} +  \frac{ 1}{  q_x}  \sum_{n,m} \frac{p_m-p_n}{\varepsilon_n-\varepsilon_m+ \omega_c} ({ \cal {I}}_i)_{nm} ([{ \cal {U}}_j,K])_{mn} \right].  \label{onsager_1}
\end{equation}
We now record the continuity equation for energy and charge. These can be compactly   written in Fourier space, for small $q$ and in the absence of external energy sources.  Using the definitions in Eq(\ref{notation_1}), we find     $[{ \cal {U}}_j,K] = q_x { \cal {I}}^\dagger_j$.  Therefore
\begin{eqnarray}
L_{ij}(q_x,\omega)& = & \frac{i}{\Omega \omega_c } \left[
- \langle [{ \cal {I}}_i,{ \cal {U}}_j] \rangle \frac{ 1}{  q_x} -    \sum_{n,m} \frac{p_m-p_n}{\varepsilon_n-\varepsilon_m+ \omega_c} ({ \cal {I}}_i)_{nm} ({ \cal {I}}^\dagger_j)_{mn} \right] .   \label{onsager_2} 
\end{eqnarray}
We next proceed to take the limit of small $q_x$. Here the inconvenient-looking first term in Eq(\ref{onsager_2}) tends to a finite limit in {\em all cases}, owing to a simple but important point. We first note that  for a large system, $K(- q_x)$ tends continuously to the Hamiltonian $K$ in the limit $q_x \rightarrow 0$. We  further note that for a generic operator $P$, the  cyclicity of trace yields  
\begin{equation}
\langle [ P, K] \rangle = \frac{1}{{\cal{ Z}}}   \mbox{Trace} \left[ \; e^{- \beta K} \; (P K -K P) \right] \equiv 0 . \label{identity_1}
\end{equation}
This relation is  noted as  the Identity-I in \cite{shastry_sumrule}.
It follows  that    $\langle [P, K(- q_x)] \rangle \propto q_x$ with a well defined coefficient Ref\cite{shastry_sumrule}. Consulting the list of  variables in 
Eq(\ref{notation_1}), we conclude that $ \lim_{q_x \rightarrow 0}  \langle [{ \cal {I}}_i,{ \cal {U}}_j] \rangle =0 $ in all cases of interest.   Observe that this result  \emph{does not require the vanishing of the commutator}  $ [ P, K]$. In the case of thermal transport $L_{22}$,  this point is important since the heat current does not commute with the Hamiltonian. In contrast, 
for $L_{11}$, i.e. electrical transport,   the charge current commutes with the total number operator and hence the limit of the ratio is well defined more trivially, leading to the familiar f-sum rule as shown below.

  In the uniform limit $ q_x \rightarrow 0$, and hence from Eq(\ref{notation_1}) we can set ${ \cal {I}}_j^\dagger = { \cal {I}}_j$. Therefore  for arbitrary frequencies, the Onsager functions read as
\begin{eqnarray}
L_{ij}(\omega)& = & \frac{i}{ \Omega \omega_c } \left[
 \langle {{\cal T}}_{ij} \rangle  -    \sum_{n,m} \frac{p_m-p_n}{\varepsilon_n-\varepsilon_m+ \omega_c} ({ \cal {I}}_i)_{nm} ({ \cal {I}}_j)_{mn}  \right],   \label{onsager_3}\\
\langle {{\cal T}}_{ij} \rangle &= & - \lim_{q_x \rightarrow 0} \langle [{ \cal {I}}_i,{ \cal {U}}_j] \rangle \frac{ 1}{  q_x} = - \lim_{q_x \rightarrow 0} \frac{ d}{d q_x} \ \langle [{ \cal {I}}_i,{ \cal {U}}_j] \rangle .   \label{gentheta} 
\end{eqnarray}
The operators ${ \cal {T}}_{ij}$ are not unique, since one can add to them  a `gauge operator'' ${ \cal {T}}^{gauge}_{ij}=[P,K]$ with arbitrary $P$,   without affecting the thermal average, due to the Identity-I \ref{identity_1} discussed above. These fundamental operators play a crucial role in the subsequent analysis, since they determine the high frequency behaviour of the response functions. These important  operators are written in a more familiar representation\cite{shastry_sumrule} as follows:
\begin{equation}
\begin{array}{|c|c|c|}
\hline
\mbox{Stress tensor} & \mbox{Thermal operator} & \mbox{Thermoelectric operator}  \\
{ \cal {T}}_{11} & { \cal {T}}_{22} &  { \cal {T}}_{12}={ \cal {T}}_{21}   \\
\tau^{xx} & \Theta^{xx}  &  \Phi^{xx}   \\
- \frac{d}{d q_x} \left[ \hat{J}_x(q_x),\rho(-q_x)\right]_{q_x \rightarrow 0}  &   - \frac{d}{d q_x} \left[ \hat{J}^Q_x(q_x),K(-q_x)\right]_{q_x \rightarrow 0} &  - \frac{d}{d q_x} \left[ \hat{J}_x(q_x), K(-q_x)\right]_{q_x \rightarrow 0} 
\label{operators}\\
 \hline
\end{array}
\end{equation}
The thermoelectric operator can also be written as 
\begin{equation}
\Phi^{xx} = { \cal {T}}_{21}=  - \frac{d}{d q_x} \left[ \hat{J}^Q_x(q_x), \rho(-q_x)\right]_{q_x \rightarrow 0},
\label{thomson}
\end{equation}
and its equivalence to the form given in Eq(\ref{operators}) amounts to showing
${ \cal {T}}_{12}={ \cal {T}}_{21}$, modulo the addition of a ``gauge operator'' discussed above. This task is more nontrivial than one might naively anticipate, and requires the use of Jacobi's identity as discussed later.

Several aspects  of Eqs(\ref{onsager_3},\ref{operators})  are worth mentioning  at this point.
\subsection{Onsager Reciprocity at finite frequencies}
 We first note that the celebrated reciprocity relations of Onsager are extended to finite $\omega$ here. These require in the present case (with no magnetic fields)
\begin{equation}
L_{ij}(\omega)= L_{ji}(\omega). \label{reciprocity}
\end{equation}
One part of the above dealing with the second term of Eq(\ref{onsager_3}) goes back to Onsager's famous argument:  in the absence of a magnetic field we can choose a real phase convention for the quantum wave functions  such that the product  $({ \cal {I}}_i)_{nm} ({ \cal {I}}_j)_{mn}$ are  real. Invariance  under complex conjugation then implies invariance under the exchange $i \leftrightarrow j$. 

The full (frequency dependent) function shows reciprocity only if we can show that ${ \cal {T}}_{ij}= { \cal {T}}_{ji}$, since this is the first part of Eq(\ref{onsager_3}).  
This identity requires the use of the Jacobi identity $0=[[a,b],c]+[[c,a],b]+[[b,c],a] $ for any three operators $a,b,c$,  and can be proved as follows. Consider ${ \cal {T}}_{12}$ which requires the first order term in $q$ of the  expectation of $ [\hat{J}_x(q),K(-q)]$. Now we use $\hat{J}_x(q) = 1/q [K, \rho(q)]$ to lowest order in $q$, so that 
\begin{eqnarray}
\langle { \cal {T}}_{12}  \rangle & = & - \left( \frac{d}{d q}\frac{1}{q} \left[ \langle [K,\rho(q)],K(-q)]\rangle \right]\right)_{q \rightarrow 0} \label{jacobi_1} \\
&=&  \left(\frac{d}{d q}\frac{1}{q} \langle \left[ [\rho(q),K(-q)], K] + [K(-q), K], \rho(q)] \right]\rangle \right)_{q \rightarrow 0} \label{jacobi_2} \\
&=&  \left( \frac{d}{d q} \langle \left[ [\hat{J}^Q_x(-q),\rho(q)] \right]\rangle \right)_{q \rightarrow 0} \label{jacobi_3}\\  
&=& \langle { \cal {T}}_{21} \rangle \label{jacobi}.
\end{eqnarray}
We used Jacobi's identity to go to Eq(\ref{jacobi_2}) from Eq(\ref{jacobi_1}), and   dropped the first term in Eq(\ref{jacobi_2}) using the Identity-I \ref{identity_1}. Eq(\ref{jacobi_3}) follows on using the definition of the heat current Eq(\ref{heat_current}). Thus we have {\em reciprocity for  all } $\omega$. A generalization to include magnetic fields can be readily made, but we skip it here.

\subsection{General Formulas for $L_{ij}(\omega)$}
We start with Eq(\ref{onsager_3}). By using a simple algebraic identity with partial fractions for arbitrary $\Delta$\cite{shastry_sumrule}, we write
$$
\frac{1}{\hbar \omega_c (\hbar  \omega_c + \Delta) } = \frac{1}{\Delta} \left( \frac{1}{\hbar \omega_c}- \frac{1}{\hbar \omega_c + \Delta} \right),
$$
we obtain
\begin{equation}
L_{ij}(\omega_c)= \frac{  i}{ \omega_c } {\cal D}_{ij}
+ \frac{  i }{  \Omega}  \sum_{n,m} \frac{ p_n- p_m }{\epsilon_m - \epsilon_n} \frac{ ({\cal I}_i)_{n m} ({\cal I}_j)_{m n} }{ \epsilon_n - \epsilon_m + \hbar \omega_c}. \label{eqkubo1}
\end{equation}
where
\begin{eqnarray}
{{\cal D}}_{ij} & = & \frac{1}{\Omega} \left[ \langle { \cal {T}}_{ij} \rangle - \sum_{n m}\frac{p_n-p_m}{\varepsilon_m - \varepsilon_n}  ({ \cal {I}}_i)_{nm} ({ \cal {I}}_j)_{mn}
\right]. \label{cald}
\end{eqnarray}
At this point it is useful to follow Kubo \cite{kubo} and introduce imaginary time operators $Q(\tau) \equiv e^{\tau K} Q e^{- \tau K}$, where $0 \leq  \tau \leq \beta$. A simple exercise in inverse Lehmann representation\footnote{Pedagogically it might be easier to go in the opposite direction, and to insert a complete set of eigenfunctions of $K$ in the expressions Eq(\ref{genkubo_1},\ref{genkubo_2}), followed by a simple integration over the imaginary time. } of the above Eqs(\ref{eqkubo1},\ref{cald}) give us the following compact Kubo type expressions\cite{kubo,shastry_sumrule} for the generalized conductivities: 
 \begin{eqnarray}
L_{ij}(\omega) & = & \frac{i}{ \omega_c } {{\cal D}}_{ij} + \frac{1}{\Omega}
\int_0^\infty \ dt \ e^{i \omega_c t} \; \int_0^\beta \; d \tau \; \langle { \cal {I}}_i( t - i \tau) { \cal {I}}_j(0) \rangle \label{genkubo_1}\\
{{\cal D}}_{ij} &=&  \frac{1}{\Omega} \left[  \langle { \cal {T}}_{ij} \rangle  -  \int_0^\beta  d \tau \langle  {\cal I}_i( - i \tau) {\cal I}_j(0) \rangle \right].  \label{genkubo_2} 
\end{eqnarray}
The stiffnesses ${{\cal D}}_{ij}$ are discussed in detail in Ref(\cite{shastry_sumrule}), and are in general non zero for all non dissipative systems such as superfluids and superconductors. For a superconductor ${\cal D}_{11}$ is the  Meissner stiffness, so that the superfluid density can be defined in terms of it\cite{shastry_sumrule}. In a superfluid or a highly pure crystal supporting second sound, the  stiffness ${\cal D}_{22}$ is non zero, and related to the  second sound  phenomenon. For dissipative systems, these stiffnesses vanish, and on dropping them from Eq(\ref{genkubo_1}) we get back the familiar Kubo type formulas\cite{kubo,luttinger}. 

\subsection{High frequency behaviour}
 The high frequency behaviour of these functions is easily found from Eq(\ref{onsager_3}) as 
\begin{equation}
\lim_{\omega \gg 0} L_{ij}(\omega) = \frac{i}{ \omega \Omega} \langle { \cal {T}}_{ij} \rangle +O(1/\omega^2).
\label{hi_1}
\end{equation}
Thus these fundamental operators determine the high frequency response, and we will pursue the consequences later.

\subsection{Sum Rules for Electrical and Thermal Conductivity}
  It is worth noting that these relations imply sum rules as well, for the thermal response functions. To see this, note that the causal nature of the Onsager coefficients and an asymptotic  fall off as inverse frequency  provides
 a dispersion relation, i.e. a Kramers Kronig relation, where $ {{\cal P}}$ represents the principal value of the integral;
\begin{eqnarray}
\Re e L_{ij}(\omega) & = & \frac{1}{\pi} {{\cal P}} \int_{- \infty}^{\infty} \frac{ d \nu}{\nu - \omega  } \Im m L_{ij}(\nu ) \label{kk_1} \\
\Im m L_{ij}(\omega) & = & \frac{1}{\pi} {{\cal P}} \int_{- \infty}^{\infty} \frac{ d \nu}{\omega - \nu} \Re e  L_{ij}(\nu). \label{kk_2}
\end{eqnarray}
We see at high frequencies  from Eq(\ref{hi_1}) and Eq(\ref{kk_2}) and   assuming the reality of the averages $\langle { \cal {T}}_{ij} \rangle $:
\begin{equation}
\lim_{\omega \gg 0} \omega \ \Im m L_{ij}(\omega) =  \frac{\langle { \cal {T}}_{ij} \rangle}{  \Omega} = \int_{-\infty}^{\infty} \ \frac{ d \nu}{\pi}  \Re e L_{ij}(\nu) \label{sumrule_1}.
\end{equation}
This relation gives all the interesting sum rules in this problem. More explicitly we find:
\begin{eqnarray}
\int_{-\infty}^{\infty} \ \frac{d \nu}{ 2 } \ \Re e  \sigma( \nu) & = & \frac{ \pi \langle \tau^{xx} \rangle }{ 2 \Omega} \label{plasma_sumrule} \\
\int_{-\infty}^{\infty} \ \frac{d \nu}{ 2} \ \Re  e  \kappa( \nu) & = & \frac{\pi \langle \Theta^{xx} \rangle}{2 T \Omega}, \label{thermal_sumrule_1}.
\end{eqnarray}
These are known as follows. (a) Eq(\ref{plasma_sumrule}) is the well known lattice plasma or f-sum rule\cite{fsumrule} with the RHS
equaling $\omega_{p}^2/8$  with $\omega_p$ as the effective plasma frequency. (b) Eq(\ref{thermal_sumrule_1}) is  the thermal sumrule\cite{shastry_sumrule} found recently. From our earlier discussion, we see that the thermal conductivity has a correction for mobile carriers Eq(\ref{thermal_zero_current}), so that we can define a finite frequency object 
\begin{equation}
\kappa_{zc}(\omega) = \frac{1}{T} \left[ L_{22}(\omega)- \frac{L_{12}(\omega)^2}{L_{11}(\omega)} \right], \label{thermal_zero_current_2}
\end{equation}
which also satisfies causality, and falls off at high frequencies as inverse $\omega$, and therefore satisfies  dispersion relations of the type Eq(\ref{kk_2}). Thus by the same argument, and using the high frequency limits of all the coefficients Eq(\ref{hi_1}),  we   infer a sum rule for this case as 
\begin{equation}
\int_{-\infty}^{\infty} \ \frac{d \nu}{ \pi}  \Re  e \kappa_{zc}( \nu)   =  \frac{1}{ T \Omega} \left[ \langle \Theta^{xx} \rangle - \frac{\langle \Phi^{xx} \rangle^2}{\langle \tau^{xx} \rangle} \right]. \label{thermal_sumrule_2}
\end{equation}
The second term in Eq(\ref{thermal_sumrule_2}) is usually small for Fermi systems at low temperatures and usually can be neglected. We may write the RHS as $\pi C_N (T) v_{eff}^2/( 2 d \Omega) $, in terms of the more conventional specific heat for a fixed number of particles, and $v_{eff}$ which is defined by this expression.
 It is interesting to note\footnote{ I thank Dr S. Mukerjee and Dr M. Peterson for interesting discussions of this point.} that the explicit dependence on the chemical potential in the RHS of  Eq(\ref{thermal_sumrule_1}) arising from the definition of $\hat{J}^Q_x$ in Eq(\ref{heat_current}), is exactly canceled in the RHS of Eq(\ref{thermal_sumrule_2}). Thus the zero current sum rule can be computed without knowing the chemical potential exactly. For immobile carriers this problem is irrelevant;  Eq(\ref{thermal_sumrule_1}) can be used without worrying about the distinction between heat current and energy current. 

We should mention that the f-sumrule Eq(\ref{plasma_sumrule}) and the thermal sum rule Eqs(\ref{thermal_sumrule_1},\ref{thermal_sumrule_2}) are both non universal in a  general system, and depend upon various material parameters and the temperature. The f-sumrule equals $\omega_p^2/8$ for quadratic bands $\varepsilon_k = \hbar^2 k^2/( 2 m)$, but in a tight binding model is related to the kinetic energy expectation. The thermal sumrule is manifestly non universal since the operators $\Theta^{xx}$ explicitly depend on the details of the Hamiltonian\cite{shastry_sumrule}. 

\subsection{Dispersion relations for Thermopower, Lorentz number and Figure of Merit}
Let us now turn to the main objects of study here namely 
\begin{eqnarray}
\mbox{Thermopower} & S(\omega)=& \frac{L_{12}(\omega)}{T L_{11}(\omega)} \nonumber \\
\mbox{Lorentz Number} & {\bf L}(\omega)= & \frac{\kappa_{zc}(\omega)}{ T \sigma(\omega)} \nonumber \\
\mbox{Figure of Merit}& {\bf Z}(\omega) T = & \frac{ S^2(\omega)}{{ \bf L}(\omega)}. \label{variables}
\end{eqnarray}
The first two objects are very   well known in transport theories Ref\cite{ziman,mahan,ashcroft_mermin}, while the figure of merit $Z T $  is a dimensionless measure of the efficacy of a thermoelectric device, with large values $Z T \sim  1$ at low $T$ being regarded as highly desirable in many applications.   Let us 
analyze these definitions and extract their dispersion relations.
It is readily seen that these variables differ  qualitatively from  the conductivity or the thermal conductivity in their high frequency behaviour. Each of these 
approaches a constant asymptotically, that can be written down by inspection.
\begin{eqnarray}
\mbox{High Freq Thermopower} & S^*= & \frac{\langle \Phi^{xx} \rangle}{T \langle \tau^{xx} \rangle } \nonumber \\
\mbox{High Freq Lorentz Number} & {\bf L}^*=& \frac{\langle \Theta^{xx} \rangle}{ T^2 \langle \tau^{xx} \rangle }- (S^*)^2 \nonumber \\
\mbox{High Freq Figure of Merit}& {\bf Z}^* T =& \frac{ \langle \Phi^{xx} \rangle^2}{\langle \Theta^{xx} \rangle \langle \tau^{xx} \rangle -\langle \Phi^{xx} \rangle^2 }. \label{hifreq_variables}
\end{eqnarray}
As a result, we can write their dispersion relations readily, they are
\begin{eqnarray}
\Re e S(\omega) & = & S^* + \frac{{\cal P}}{\pi}\int_{-\infty}^\infty \frac{ d \ \nu}{\nu - \omega   } \; \Im m S(\nu) \label{dis_s}\\
\Re e {\bf L}(\omega) & = & {\bf L}^* + \frac{{\cal P}}{\pi}\int_{-\infty}^\infty \frac{ d \ \nu}{\nu - \omega   } \; \Im m {\bf L}(\nu) \label{dis_l}\\
\Re e {\bf Z}(\omega) & = & {\bf Z}^* + \frac{{\cal P}}{\pi}\int_{-\infty}^\infty \frac{ d \ \nu}{\nu - \omega   } \; \Im m {\bf Z}(\nu). \label{dis_z}
\end{eqnarray}
These transport quantities are generally real at only two values of frequency, namely zero or infinity, and are very similar in mathematical structure to the Hall resistivity discussed in Eq(\ref{hall_dispersion}). The imaginary part is expected to go linearly at small $\omega$,  falling off over some finite interval in $\omega$ corresponding to the energy range of the contributing physical processes. Thus the difference between the DC transport and high frequency values can be expressed in all these cases as an integral over the imaginary part of these three variables divided by the frequency, and may be amenable to direct  measurements, as in the case of the Hall effect. 

\section{Thermoelectric Phenomena in Correlated Matter}
\subsection{Limiting case of free electrons, $S^*$  the Heikes Mott and Mott results}
We propose  the use of the high frequency variables Eq(\ref{hifreq_variables}) in correlated matter, for reasons that are essentially the same as those for proposing the high frequency Hall constant, explained earlier. These variables are singled out by the fact that they have a finite limit at high $\omega$, as compared to say $\kappa(\omega)$ or $L_{12}(\omega)$, which vanish in that limit.
 In particular we expect that these high frequency limits of the three variables listed in Eq(\ref{hifreq_variables}), are good indicators of the DC transport measurements in correlated matter, where we can use the projected $t$-$J$ model, whereas for the Hubbard model, these should be good only for intermediate to weak coupling. The origin of this expectation is not repeated here since it is identical to  the argument given for the Hall constant after Eq(\ref{tJ1}) and the later paragraphs. In the following, we will see the consequences of this proposal, and estimate its accuracy in some well controlled examples. By way of motivating this calculation, we show in Fig(\ref{hall_thermopower}) the computed Hall and Seebeck coefficients for the triangular lattice, where these objects have a similar behaviour as a function of electron filling in a Mott Hubbard system.

Let us begin by    listing the three basic operators for the simplest Drude Sommerfeld type model of a free  electron gas,  with particle scattering off some impurities or phonons characterized by a relaxation time $\tau$. Let the  particle energy dispersion be denoted by  $\varepsilon_k$, and their group velocity $v_p^x = \partial \varepsilon_k/ \partial k_x$.  A small calculation of Eq({\ref{operators}) shows 
\begin{eqnarray}
\tau^{xx} & = & q_e^2 \sum_{p,\sigma} \frac{\partial}{\partial p_x} \left\{ v_p^x \right\}\;c^\dagger_{p,\sigma} c_{p,\sigma} \nonumber \\
\Theta^{xx} & = &  \sum_{p,\sigma}\frac{\partial}{\partial p_x} \left\{ v_p^x  (\varepsilon_p - \mu)^2 \right\} c^\dagger_{p,\sigma} c_{p,\sigma}  \nonumber , \\
\Phi^{xx} & = & q_e \sum_{p,\sigma}\frac{\partial}{\partial p_x} \left\{ v_p^x  (\varepsilon_p - \mu)\right\} c^\dagger_{p,\sigma} c_{p,\sigma} \label{free_ops}
\end{eqnarray}
 We next form the thermal  averages,
\begin{eqnarray}
\langle \tau^{xx} \rangle &=& {2 q_e^2} \sum_p \;\; n_{p}\;\;   \frac{d}{d p_x} 
\left[  v^x_{p} \right] \nonumber \\ 
\langle \Theta^{xx} \rangle &=& {2} \sum_p \;\; n_{p}\;\; \frac{d}{d p_x} 
\left[  v^x_{p} (\varepsilon_{p} - \mu)^2 \right]
  \nonumber\\
\langle \Phi^{xx} \rangle &=& {2 q_e} \sum_p \;\; n_{p}\;\; \frac{d}{d p_x} 
\left[  v^x_{p} (\varepsilon_{p} - \mu) \right].
\end{eqnarray}
Here $n_{p}$ is the Fermi function. we now focus on the low $T$ behaviour of these formulae.  At low temperatures $T$, we use the Sommerfeld expansion\cite{ashcroft_mermin} after integrating by parts,  and  obtain the leading   behaviour:
\begin{eqnarray}
\langle \tau^{xx} \rangle &=& \Omega \ 2 \ q_e^2 \rho_0(\mu) \ \langle (v^x_{p})^2\rangle_{\mu} \nonumber \\
\langle \Theta^{xx} \rangle &=& {\Omega}\ T^2 \ \frac{2 \pi^2 k_B^2 }{3} \ \rho_0(\mu) \ \langle (v^x_{p})^2\rangle_{\mu}
 \nonumber \\
\langle \Phi^{xx} \rangle &=& {\Omega} \ T^2 \ \frac{2 q_e \pi^2 k_B^2 }{3} \ \left[ 
\rho_0'(\mu) \langle (v^x_{p})^2\rangle_{\mu} +\rho_0(\mu) \frac{d}{d \mu} \langle (v^x_{p})^2\rangle_{\mu}
\right], \label{freetheta}
\end{eqnarray}
where $\rho_0(\mu)$ is the density of states per spin per site at the Fermi level $\mu$ and the primes denote derivatives w.r.t. $\mu$,  the average is over the Fermi surface as usual.  We may form the high frequency ratios as in Eq(\ref{hifreq_variables}), and get the leading formulas\footnote{The reader is requested to ignore the irksome  issue of the dimensionality of the argument of the logarithm.  The logarithm is just a notational  device to collect the coefficients in this formula and in Eqs(\ref{mott},\ref{kelvin_5}).} 
\begin{eqnarray}
S^* &=& T \frac{ \pi^2 k_B^2 }{ 3 q_e} \frac{d}{d \mu} \ln \left[ \rho_0(\mu) \langle (v^x_{p})^2\rangle_{\mu} \right]  \label{sstar_free} \\
L^* &=& \frac{ \pi^2 k_B^2}{ 3 q_e^2}. \label{lstar_free}
\end{eqnarray}
These formulas are indeed very close to what we expect from the Bloch Boltzmann theory.
 The high frequency result gives {\em the same}  Lorentz number as we get from  the Bloch Boltzmann theory.
In the Bloch Boltzmann theory, the thermopower can be calculated assuming an energy momentum dependent relaxation time $\tau(p,\omega)$,   as
\begin{equation}
S_{Mott}   =  T \frac{ \pi^2 k_B^2 }{ 3 q_e} \frac{d}{d \mu} \ln \left[ \rho_0(\mu) \langle (v^x_{p})^2 \tau(p,\mu)\rangle_{\mu}  \right]  , \label{mott}
\end{equation}
and is referred to as the Mott  formula\cite{ashcroft_mermin,ziman}.
A comparison between the two formulae Eq(\ref{sstar_free}, \ref{mott}) for the thermopower reveals the nature of the high frequency limit: it ignores the energy dependence of the relaxation time, but captures the density of states. Thus this formalism is expected to be accurate whenever the scattering is less important than say the density of states and  correlations.

If the free electron gas in the above discussion is replaced by electrons that interact with each other, in addition to scattering off impurities or phonons or amongst themselves, the details of the interactions become crucial in writing the operators analogous to Eq(\ref{free_ops})down.
The thermal   operators $\Theta^{xx}$ can be computed for any given model by a prescription set out in \cite{shastry_sumrule}, and detailed expressions are available there for many standard electronic models:  the Hubbard model, the $t$-$J$ model, and the Anderson model. Also corresponding expressions are available heat conduction in insulators such as the Heisenberg antiferromagnet, and  for non linear lattice models such as the Fermi Pasta Ulam chain \cite{fpu}. The thermoelectric operators $\Phi^{xx}$ are also given explicitly for  the conducting models for the same set of models in the same reference. Given their length it seems hardly worthwhile to reproduce them here. We merely mention that the operators involve the interaction parameters, just as the energy currents do, and have to be worked out for each model individually. The one exception is the $\tau^{xx}$ operator, which usually  has  the same form as in Eq(\ref{free_ops}), due to the fact that interactions are velocity independent. We will see the explicit form of the $\Phi^{xx}$ operator for the $U= \infty$ Hubbard model below Eq(\ref{phi}).

Let us also note the general formula for the thermopower from Eqs(\ref{variables},\ref{genkubo_1}). On dropping the second term Eq(\ref{genkubo_2}), we get the standard formulas appropriate for dissipative systems, where we can write the ``exact'' Kubo formula\cite{kubo}:
\begin{equation}
S_{Kubo} = \left[ \frac{\int_0^\infty \ dt  \; \int_0^\beta \; d \tau \; \langle \hat{J}^E_x( t - i \tau) \hat{J}_x(0) \rangle }{\int_0^\infty \ dt  \; \int_0^\beta \; d \tau \; \langle \hat{J}_x( t - i \tau) \hat{J}_x(0) \rangle} \ -\frac{\mu(0)}{q_e} \right]+ \frac{\mu(0) - \mu(T)}{q_e}  \label{kubo_1}.
\end{equation}
We have used Eq(\ref{heat_current_decomposition}) and further added and subtracted the
$\frac{\mu(0)}{q_e} $ term for convenience, to arrive at Eq(\ref{kubo_1}). The Mott result Eq(\ref{mott}) follows from this general formula in the limit of weak scattering, as textbooks indicate\cite{mahan}. For narrow band systems, Heikes introduced another  approximation popularized by Mott\cite{beni,heikes}, namely the Heikes Mott formula
\begin{equation}
S_{HM} =  \frac{\mu(0) - \mu(T)}{q_e}. 
\label{mott_heikes}
\end{equation}
This formula emphasizes the thermodynamic interpretation of the thermopower, this term can be loosely regarded as the entropy per particle\footnote{Strictly speaking   $\mu$ is  a derivative of the entropy w.r.t. the number of particles, i.e.  $\mu(T)= - T (\partial S(N,T) /\partial N)_{E,T}$.}. This motivates us\cite{shastry_sumrule} to decompose the thermopower as
\begin{equation}
S_{Kubo}= S_{Tr}+ S_{HM}, \label{kubo_2}
\end{equation}
thereby defining the ``transport part'' of thermopower as the first part of Eq(\ref{kubo_1}) evaluated with $\mu(0)$, as opposed to the thermodynamic part $S_{HM}$.
Using the high frequency approximation Eq(\ref{hifreq_variables}), we approximate (only) the transport part in Eq(\ref{kubo_2}) and  write
\begin{eqnarray}
S^* & = &  S^*_{Tr}+ S_{HM}, \nonumber \\
S^*_{Tr} & = & \frac{\langle \Phi'^{xx} \rangle}{ T \langle \tau^{xx}\rangle } , \label{hifreq_s}
\end{eqnarray}
with the $\Phi'^{xx}$ differs from   the variable $\Phi^{xx}$ in that the chemical potential $\mu$  is replaced by  the $T=0$ value  $\mu(0)$. The low $T$ limit for the free particle case of this relation   is given in  Eq(\ref{sstar_free}). For a correlated many body system, it is much easier to work with this variable. The computational advantage in Eq(\ref{hifreq_s}) over Eq(\ref{kubo_1}) is that the transport part is approximated by an equal time correlator as opposed to a dynamical correlator. This allows us to apply one of several possible techniques to the problem, such as exact diagonalization and also high $T$ expansions.

\subsection{Kelvin's thermodynamical formula for thermopower}

It is  interesting  to discuss  Kelvin's thermodynamic derivation of the thermopower \cite{kelvin}.  In his seminal work, Onsager\cite{onsager} discussed   Kelvin's derivation of reciprocity   given several decades earlier. He argued  that the phenomenon of transport, including reciprocity, cannot be understood within equilibrium thermodynamics or statistical mechanics. Interestingly as late as 1966, Wannier wrote in his  textbook\cite{wannier}: ``{\em Opinions are   divided as to whether Kelvin's derivation is fundamentally flawed or not}''.  A detailed account of this debate and its resolution seem to be missing in  literature. 

Our discussion of the thermopower takes us to the brink of this old debate, and so we make a small excursion to obtain a thermodynamic approximation of the correct answer.  This derivation  captures the spirit of the Kelvin argument, and  provides an approximate expression for the thermopower $S$.  For deriving this, let us rewind to Eq(\ref{definitionL}) of the finite  ${q},\omega$ dependent Onsager coefficients $L_{ij}(q,\omega)$. Using Eqs(\ref{notation_1},\ref{response1},\ref{chidef}) we see that
\begin{eqnarray}
 L_{11}(q,\omega)& = & \frac{i}{\Omega q_x} \chi_{{\hat{J}}_x(q_x), \rho(-q_x)}(\omega) \nonumber \\
L_{12}(q,\omega)& = & \frac{i}{\Omega q_x} \chi_{{\hat{J}}_x(q_x), K(-q_x)}(\omega), \;\;\;\mbox{hence}\nonumber \\
 S(q_x,\omega) & = & \frac{ \chi_{{\hat{J}}_x(q_x), K(-q_x)}(\omega)}{T \ \chi_{{\hat{J}}_x(q_x), \rho(-q_x)}(\omega)} \label{kelvin_1}.
\end{eqnarray}
Onsager's prescription at this point is to take the transport limit, i.e. first let $q_x \rightarrow 0$ followed by the static limit, to get the exact formula\cite{luttinger,kubo}. We saw in the previous section that this ratio has another finite and interesting limit, leading to $S^*$, when we let $q_x \rightarrow 0$ followed by $\omega \gg 0$. It is interesting and amusing  that in the opposite slow limit, i.e. 
$\omega \rightarrow 0$ followed by $q_x \rightarrow 0$, once again $S(q_x ,\omega)$ has a finite and well defined result. This limit is what we identify with the Kelvin calculation and his formula, since the objects that arise are purely equilibrium quantities. Thus
\begin{eqnarray}
S_{Kelvin} & = & \lim_{q_x \rightarrow 0, \omega \rightarrow 0}S(q_x,\omega) \nonumber \\
S(q_x,\omega) &=& \frac{ \chi_{[K, \rho(q_x)], K(-q_x)}(\omega)}{T \ \chi_{[K, \rho(q_x)], \rho(-q_x)}(\omega)} \label{kelvin_2} \\
&=& \frac{ \chi_{ \rho(q_x), K(-q_x)}(\omega)}{T \ \chi_{ \rho(q_x), \rho(-q_x)}(\omega)}. \label{kelvin_3} 
\end{eqnarray} 
We have used the continuity relation $\hat{J}_x(q_x) = \frac{1}{q_x} [K, \rho(q_x )]$ to go from Eq(\ref{kelvin_1}) to Eq(\ref{kelvin_2}). The next stage involves writing
a Ward type identity, 
\begin{equation} \chi_{[K, \rho(q_x)], K(-q_x)}(\omega) = - \omega \chi_{ \rho(q_x), 
K(-q_x)}(\omega) + \langle [K(-q_x),\rho(q_x)] \rangle, \label{ward_1} 
\end{equation}  
and a similar one for the denominator, followed by realizing that the second term of the r.h.s. of Eq(\ref{ward_1}) vanishes on using parity for any finite $q$ in a system with inversion symmetry\footnote{ The argument is trivial for the denominator since density fluctuations commute at different wave vectors. In the numerator, consider the real expectation $\gamma(q_x) = \langle [K(-q_x),\rho(q_x)] \rangle$. Clearly $\gamma^*(q_x)  = \langle [\rho(-q_x), K(q_x)] \rangle = - \gamma(-q_x)$. But from parity $\gamma(-q_x)= \gamma(q_x)$ and hence the result $\gamma(q_x)=0$.}. We can now take the static limit and get the equilibrium Kelvin result
\begin{equation}
S_{Kelvin} = \lim_{q_x \rightarrow 0} \frac{ \chi_{ \rho(q_x), K(-q_x)}(0)}{T \ \chi_{ \rho(q_x), \rho(-q_x)}(0)}, \label{kelvin_4}
\end{equation}
where in the limit,  the denominator is related to the  thermodynamic compressibility, and the 
numerator is an equilibrium cross correlation function  between energy and charge density. It is  straightforward to see that reciprocity  holds in this sequence of limits as well.

In the case of free particles, it is easy to evaluate Eq(\ref{kelvin_4}) at low $T$  and we find
\begin{equation}
S_{Kelvin}=  T \frac{ \pi^2 k_B^2 }{ 3 q_e} \frac{d}{d \mu} \ln \left[ \rho_0(\mu)  \right]. \label{kelvin_5} 
\end{equation} 
It is amusing to compare Eq(\ref{sstar_free},\ref{mott}) and Eq(\ref{kelvin_5}). Compared to the ``exact'' Mott formula that follows from the Onsager limiting procedure, $S^*$ captures the  answer except for  the energy dependent relaxation rate. The Kelvin formula further approximates $S^*$ by neglecting the energy dependence of the velocity average. Thus we conclude that the Kelvin approximation is  inferior to the high frequency approximation, but does capture the density of states effects. 

 The above,  rather formal manipulation with the limits, can be  nicely visualized by working instead with open boundary conditions. Let us imagine a ``gedanken experiment'', where a long   isolated cylinder of the material of interest, with length L, is subjected to a time varying temperature gradient. Since this experiment is exempt from practical issues, we further  imagine a Luttinger version of this, where a pair of tiny blackholes \footnote{ If the earth became a black hole it would have a diameter of about 0.017 meters, about the size of a marble. {\url{ http://www.windows.ucar.edu/tour/link=/kids_space/black.html&edu=elem}}
} are brought to the two ends of the sample (somehow!!), and oscillated in space and time. Thus  we  apply a space-time varying gravitational field $\psi(\vec{x},t) = \delta \psi_0 \ \frac{x}{L} \exp \{- i \omega t\} $ together with a similar electrostatic potential $\phi(\vec{x},t)$,  and compute the induced oscillating dipole moment $P= \sum_x  x \rho(\vec{x})$ using perturbation theory. The gravitational field is again a proxy for temperature variation. By forming the ratio of the gravitational field amplitude $\delta\psi_0$ to the electrostatic amplitude $\delta\phi_0$
needed to produce a given dipole moment, we can extract the  thermopower. The rigorously correct transport limit, as  applied to  this situation,  requires {\em the thermodynamic limit to be taken before} $\omega \rightarrow 0$.  If we compute the opposite limit instead, i.e. a finite system and a DC field, then the result maps to the above Eq(\ref{kelvin_4}). Such a limiting process is tempting from the physical picture of the so called ``absolute thermopower''.  In this case,  one studies  a single system  with applied thermal gradients, which  develops a voltage across its ends.   This type of a  picture was presumably behind the Kelvin derivation.

\subsection{Applications to Sodium Cobaltates in the Curie Weiss Metallic Phase}
At this point it is worthwhile to compare the results of various approximations in the important and current problem of sodium cobaltates $Na_x Co O_2$, with $x\sim .68$. Recent interest in this system started with the observation of high thermopower ($S\sim 80 \mu V/K$) at room temperatures in this system by Terasaki\cite{terasaki}. Wang, Rogado, Cava and Ong, in another important paper\cite{ong}  found that this thermopower is strongly magnetic field dependent. They  further found that the metallic conduction  is coexistent with a Curie Weiss susceptibility characteristic of insulators. This has given rise to the nomenclature of a Curie Weiss metallic phase. The basic modeling of this system, as suggested by Wang {\em et. al.}, is in terms of a strongly correlated Fermi system, with  no double occupancy of holes. The holes move on a triangular lattice provided by the $Co$ atoms, and the system may be regarded, to a first approximation, as a bunch of uncoupled 2D triangular lattice planes with a $t$-$J$ model description of correlated holes. After performing a particle hole transformation we can write the basic Hamiltonian as 
\begin{equation}
H= - \sum_{\vec{x},\vec{\eta}} t(\ven) \tilde{c}^\dagger_{\vec{x} + \vec{\eta},\sigma} \tilde{c}_{\vec{x},\sigma} + J \sum_{<ij>} \vec{S}_i . \vec{S}_j .\label{tjham}
\end{equation}
Here $\vec{\eta}$ is the nearest neighbor vector on the triangular lattice.
\begin{figure}[h]
\includegraphics[width=9cm]{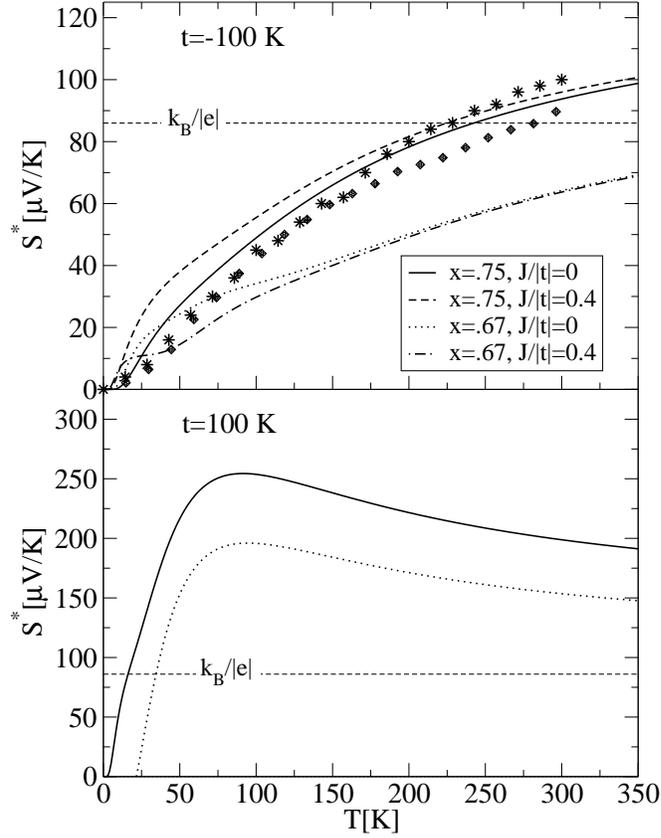}
  \caption{{\bf Upper Panel:}  Thermopower computed  for the triangular lattice $t$-$J$ model in Ref\cite{haerter_peterson_shastry_curie_weiss},  compared to the experimental data of Ref\cite{terasaki} (stars)  and Ref\cite{ong}(diamonds). The absolute scale is set by a single parameter $t=-100^0$K. The different curves correspond to various values of  doping $x$ and  $J/|t|$. {\bf  Lower Panel:}  This shows the effect of reversing the sign of hopping in this system. This is  a prediction of this theory for a fiduciary hole doped sodium cobaltate type system. The peak value of $250 \mu V/K$ can be further manipulated upwards by changing material parameters $J ,x$.}
\label{sstar_vs_T}
\end{figure}
This model corresponds to  the limit of $U\rightarrow \infty$. In this limit the Fermionic commutation relations need to be modified into the Gutzwiller-Hubbard  projected operator\cite{hubbardops} relations (with $\bar{\sigma}= - \sigma $) 
\begin{eqnarray}
\tilde{c}_{\vec{x},\sigma}  &=&  P_G  \; c_{\vec{x},\sigma} \; P_G   \nonumber \\
\left\{\tilde{c}_{\vec{x},\sigma}, \tilde{c}^\dagger_{\vec{x'},\sigma'} \right\} &=&  \delta_{\vec{x},\vec{x'}}  \left\{ \delta_{\sigma,\sigma'} ( 1 - n_{\vec{x},\bar{\sigma}})  + ( 1 - {\delta_{\bar{\sigma},\sigma'}} ) \tilde{c}^\dagger_{\vec{x},\sigma} \tilde{c}_{\vec{x},\bar{\sigma} } \right\} \nonumber\\  
&  \equiv &Y_{\sigma,\sigma'} \; \delta_{\vec{x},\vec{x'}}
\end{eqnarray}
The presence of the $Y$ factor is due to strong correlations, and makes the computation nontrivial. The number operator $n_{\vec{x},\sigma}$ is unaffected by the projection. 
Let us consider the kinetic energy only, i.e. the $t$ part, since this  is expected to dominate in transport properties, at least far enough from half filling and for $t \gg J$. The addition of the $J$ part can be done without too much difficulty, in fact  the numerics discussed below include the full Hamiltonian.

Let us note down the expressions for the charge current and the energy current at finite wave vectors by direct computation:
\begin{eqnarray}
\hat{K}(k) &=& - \sum_{\vec{x},\vec{\eta},\sigma} ( t(\vec{\eta})+ \mu \delta_{\vec{\eta},0}) \; e^{ i \vec{k}.(\vec{x}+\frac{1}{2} \vec{\eta})}\; \tilde{c}^\dagger_{\vec{x}+\vec{\eta},\sigma} \tilde{c}_{\vec{x},\sigma} \label{ops_tj} \\
\hat{J}_x(k) &=&  i q_e \sum_{\vec{x},\vec{\eta},\sigma} \;\eta_x t(\vec{\eta})   \; e^{ i \vec{k}.(\vec{x}+\frac{1}{2} \vec{\eta})} \; \tilde{c}^\dagger_{\vec{x}+\vec{\eta},\sigma} \tilde{c}_{\vec{x},\sigma} \nonumber \\
\hat{J}^Q_x(k)  &=& -\frac{i}{2 }\sum_{\vec{x},\vec{\eta},\vec{\eta}',\sigma}\;( \eta_x + \eta'_x)  t(\vec{\eta}) t(\vec{\eta}')  \; e^{ i \vec{k}.(\vec{x}+\frac{1}{2} (\vec{\eta}+\vec{\eta}'))} \; Y_{\sigma',\sigma}(\vec{x}+\vec{\eta'})\;\tilde{c}^\dagger_{\vec{x}+\vec{\eta}+\vec{\eta'},\sigma'} \tilde{c}_{\vec{x},\sigma} \nonumber \\
&& - \frac{ \mu}{q_e} \hat{J}_x(k) \nonumber
\end{eqnarray}
We evaluate the thermoelectric  operator as:
\begin{eqnarray}
 \Phi^{xx} & = &  -\frac{q_e}{2} \sum_{\vec{\eta},\vec{\eta',\sigma,\sigma'},\vec{x}} ( \eta_x+\eta'_x)^2 \; t(\vec{\eta})\; t(\vec{\eta'}) \; Y_{\sigma',\sigma}(\vec{x}+\vec{\eta}) \; \tilde{c}^\dagger_{\vec{x}+\vec{\eta}+\vec{\eta'},\sigma'} \tilde{c}_{\vec{x},\sigma}
\nonumber \\
&&  - q_e \mu \sum_{\vec{\eta},\sigma,\vec{x}} \; \eta_x^2 \; t(\vec{\eta}) \; \tilde{c}^\dagger_{\vec{x}+\vec{\eta},\sigma} \tilde{c}_{\vec{x},\sigma}. \label{phi} 
\end{eqnarray}
This expression gives an idea of the complexity of the operators that arise in the theory. Let us first present some numerical results obtained by exact diagonalization\cite{peterson_1,peterson_2} of small clusters of the triangular lattice. We can compute all eigenstates and matrix elements for up to 14 or 15 site clusters of the triangular lattice. We can   therefore assemble the full dynamical conductivities from Eq(\ref{variables}). The involved calculations are fully described in the papers\cite{peterson_1,peterson_2}, and we will content ourselves with displaying the main results. Firstly, consider the absolute scale of the thermopower $S^*$ as a function of temperature, shown in Fig(\ref{sstar_vs_T}).
The upper panel in Fig({\ref{sstar_vs_T}) shows that  this comparison with experiment  is quite successful on a quantitative scale.  One can next ask, how good is the 
approximation of infinite frequency, purely in theoretical terms. To answer this we compute the frequency dependence of $S(\omega)$, as shown in Fig(\ref{sstar_vs_w}). It is clear from this figure that the approximation of high frequency is  excellent, the maximum error being less than $3 \%$. Thus we are computing essentially the DC transport object, at least for clusters of these sizes.
This benchmarking gives us confidence in the results of the  high frequency formulas for  thermopower.
\begin{flushleft}
\begin{figure}[h]
\includegraphics[width=7.5cm, angle = -90]{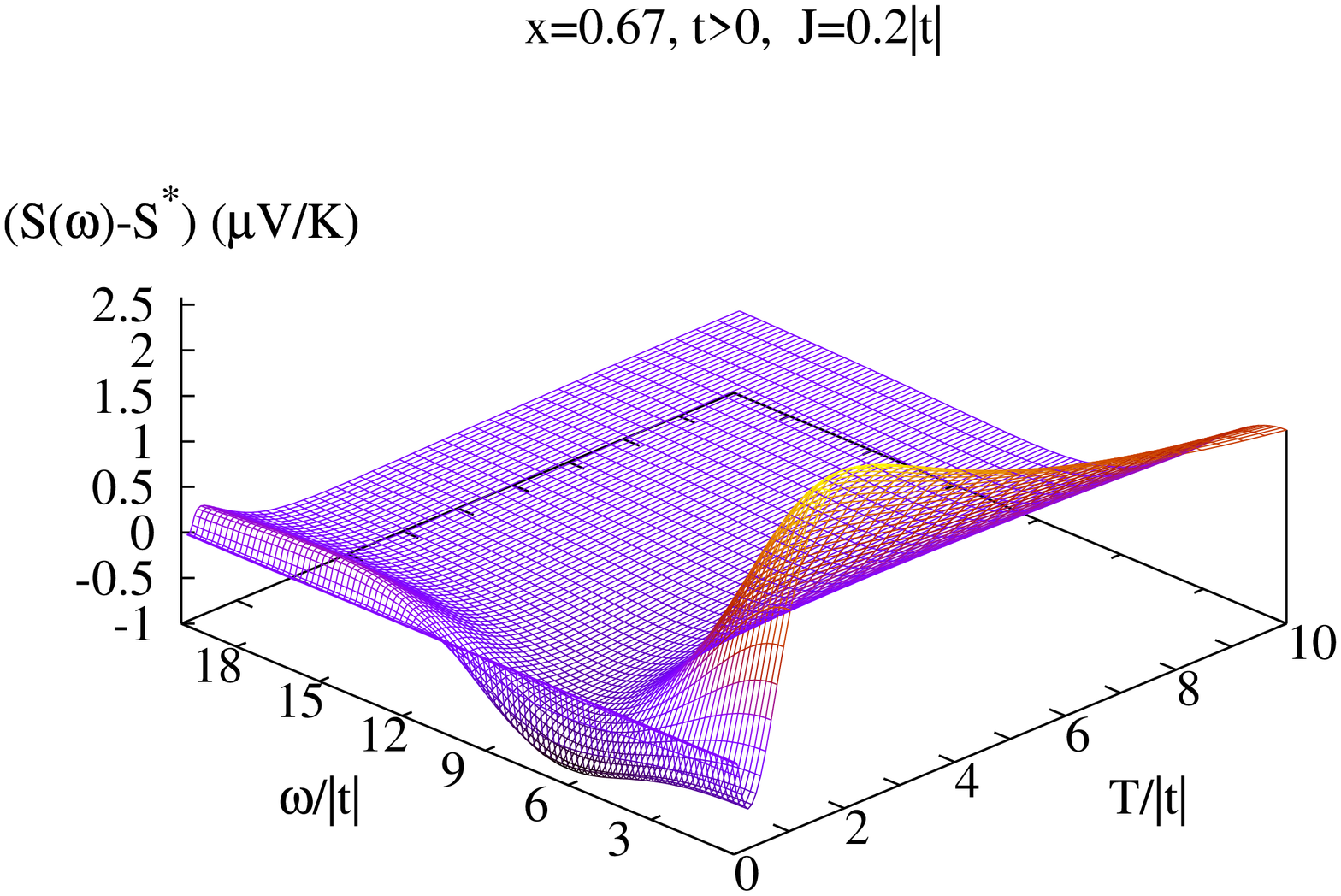}
  \caption{The frequency dependence of the thermopower computed   for the triangular lattice $t$-$J$ model from Ref \cite{peterson_2}.   The sign of hopping is flipped relative to that in Fig(\ref{sstar_vs_T}) in this and all other figures by using a p-h mapping. Recalling that the scale of $S\sim 100 \mu V/K$, we conclude that the frequency dependence is indeed very small ($\sim 3 \%$ at most). }
\label{sstar_vs_w}
\end{figure}
\end{flushleft}

\subsection{High temperature expansion for Thermopower}

We next show that rather simple considerations of our formulae lead to an important prediction for enhancing the thermopower for a triangular lattice system with a suitable choice of the hopping parameter. We find a  remarkable   effect of the sign of hopping on the transport part of the thermopower. This is well illustrated in   the lower panel of Fig(\ref{sstar_vs_T}). This  shows the  enhancement of the computed thermopower at low and intermediate $T$'s, achieved by flipping the sign of hopping from the upper panel. We perform a simple computation at high $T$ that throws light on this phenomenon. We focus on the kinetic energy which is expected to dominate the transport contributions. Let us compute the thermopower $S^*$ from Eqs(\ref{phi},\ref{hifreq_variables})
\begin{equation}
S^*= - \frac{\mu}{q_e T}+ \frac{ q_e \Delta}{ T \langle \tau^{xx} \rangle} \label{s1}
\end{equation}
where
\begin{equation}
\Delta=  -\frac{1}{2 } \sum_{\vec{\eta},\vec{\eta'},\vec{x}} ( \eta_x+\eta'_x)^2 \; t(\vec{\eta})\; t(\vec{\eta'}) \; \langle Y_{\sigma',\sigma}(\vec{x}+\vec{\eta}) \;  \tilde{c}^\dagger_{\vec{x}+\vec{\eta}+\vec{\eta'},\sigma'} \tilde{c}_{\vec{x},\sigma} \rangle \label{Delta}
\end{equation}
The computation of the different parts proceeds as follows: we show readily that (for the hole doped case) using translation invariance and with $n$ as the number of particles per site at high T,
\begin{equation}
 \langle \tau^{xx} \rangle = 6 {\Omega} q_e^2 t \langle \tilde{c}^\dagger_1 \tilde{c}_0 \rangle \sim 3 {\Omega} q_e^2 \beta t^2 n(1-n).
\end{equation}
The structure of the term Eq(\ref{Delta}) is most instructive. At high temperatures, for a square lattice we need to go to second order in $\beta t$ to get a contribution  with $\eta_x+\eta'_x \neq 0$, to the expectation of the hopping $\langle \tilde{c}^\dagger_{\vec{x}+\vec{\eta}+\vec{\eta'},\sigma'} \tilde{c}_{\vec{x},\sigma} \rangle$. For the triangular lattice, on the other hand,  we already have a contribution at  first order.  For the triangular lattice, corresponding to each nearest neighbor, there are precisely two neighbors where the third hop is a nearest neighbor hop. A short calculation gives
\begin{equation}
\Delta \sim  -  3 {\Omega} t^2 \sum_{\sigma,\sigma'} \langle Y_{\sigma',\sigma}(\vec{\eta}) \tilde{c}^\dagger_{\vec{\eta}+\vec{\eta'},\sigma'} \tilde{c}_{\vec{0},\sigma} \rangle. 
\end{equation}
The spins must be the same to the leading order in $\beta t$ where we  generate a hopping term $\tilde{c}^\dagger_{\vec{0},\sigma}\tilde{c}_{\vec{\eta}+\vec{\eta'},\sigma} $ from an expansion of $\exp(-\beta K)$, and hence a simple estimation yields
\begin{equation}
\Delta = - \frac{3}{2 } {\Omega} t^3 \beta n(1-n)(2-n) +O(\beta^3).
\end{equation}
This together with $\mu/ k_B T = \log(n/2(1-n)) + O(\beta^2 t^2)$ gives us the result for $ 0\leq n \leq 1$
\begin{equation}
S^*= \frac{k_B}{q_e} \left\{ \log[2(1-n)/n] - \beta t \frac{2-n}{2} + O(\beta^2 t^2) \right\}, \label{s_holedoping}
\end{equation}
and 
\begin{equation}
S^*= - \frac{k_B}{q_e} \left\{  \log[2(n-1)/(2-n)] + \beta t \frac{n}{2} + O(\beta ^2 t^2) \right\} \label{s_electrondoping}
\end{equation}
for $1 \leq n \leq 2$ using particle hole symmetry\cite{shastry_sumrule}.
\begin{flushleft}
\begin{figure}[h]
\includegraphics[width=9cm, angle = -90]{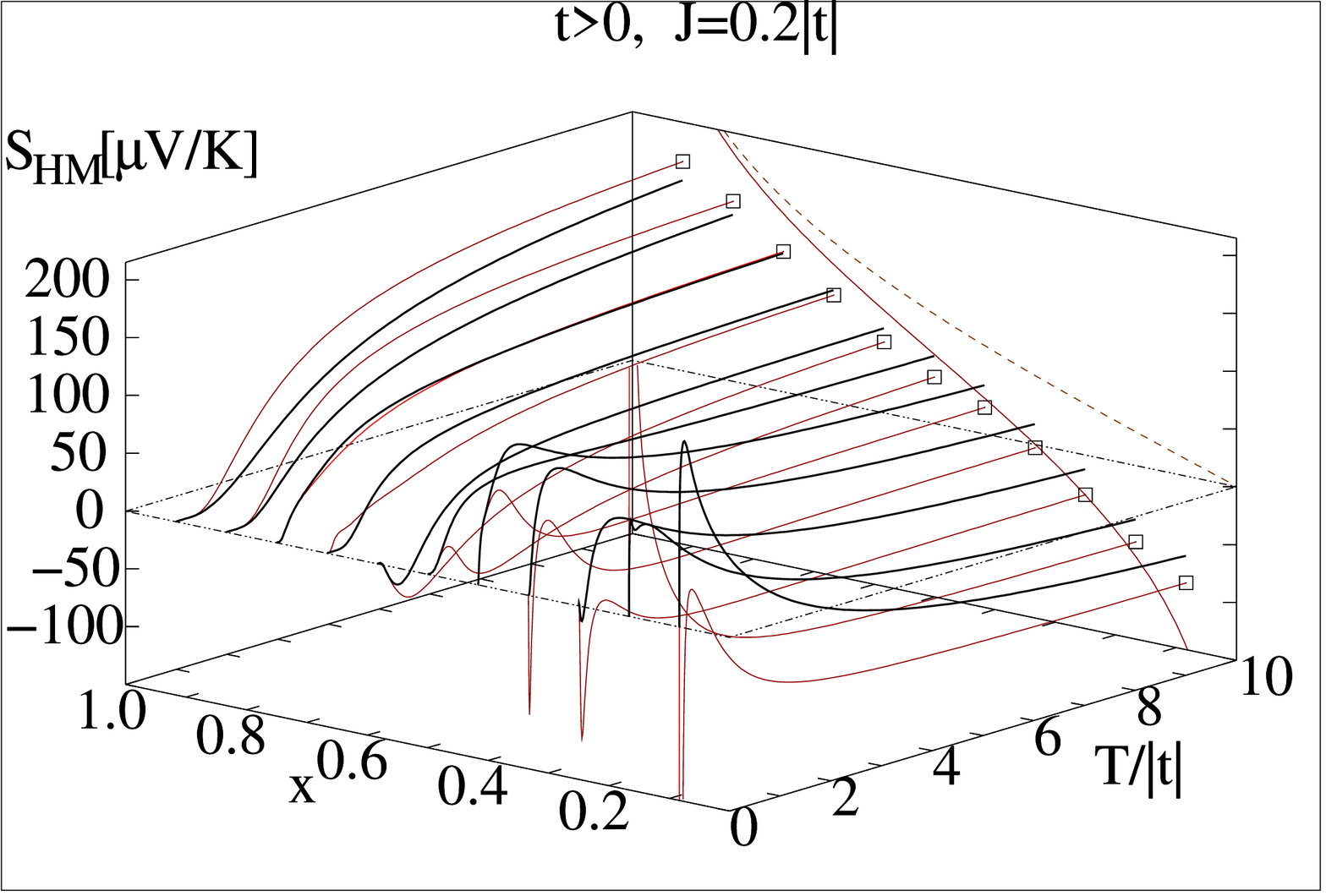}
\includegraphics[width=9cm, angle=-90]{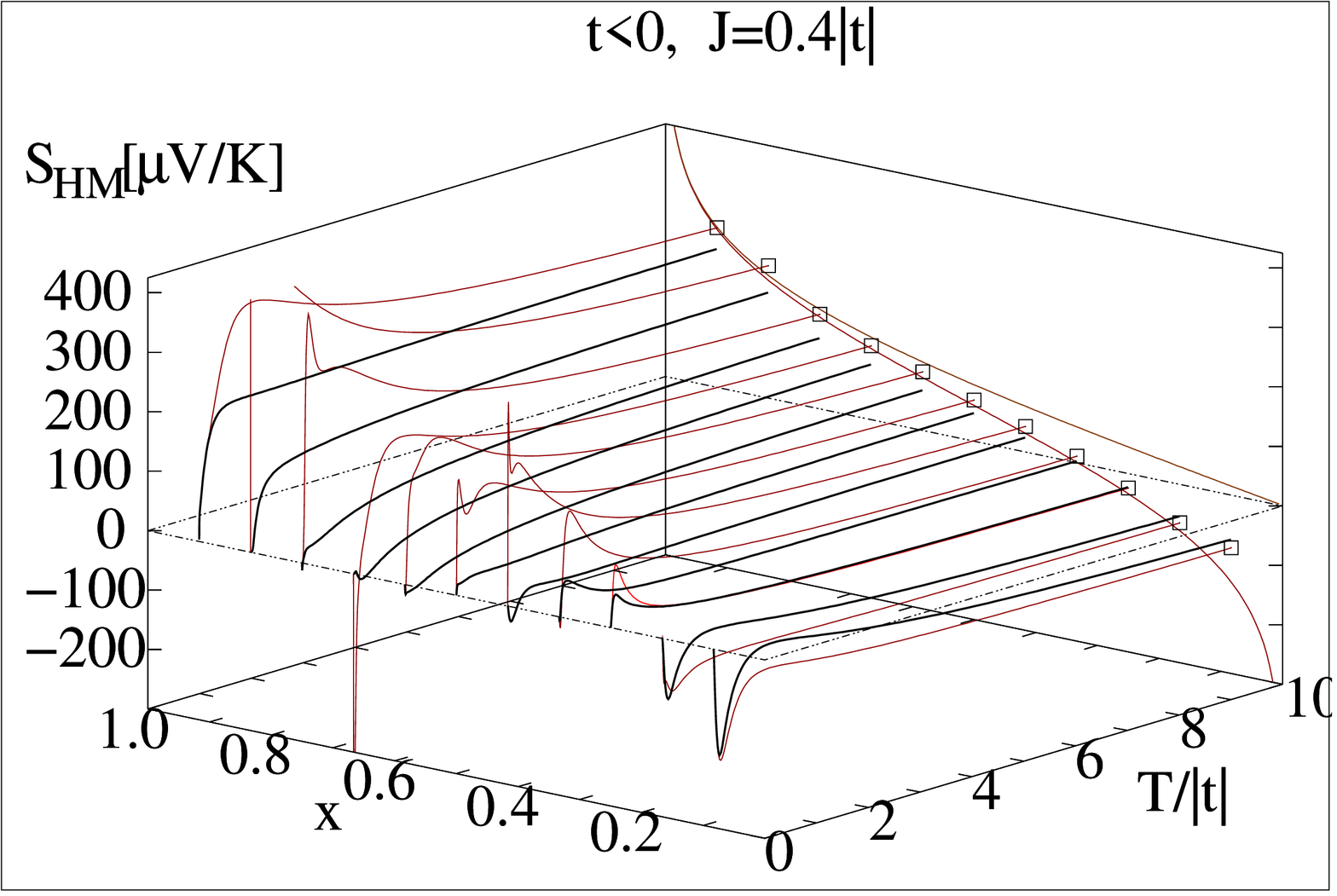}
\caption{The thermopower versus filling $x=1-n$ and temperature $T$ in the $t$-$J$ model from numerical studies\cite{peterson_1,peterson_2} on clusters of the  triangular lattice. In both cases the lower curves correspond to the Heikes Mott formula Eq(\ref{mott_heikes}) and upper to the high frequency result of Eq(\ref{hifreq_variables}). [{\bf Top:}]  The case of the  sodium cobaltates, i.e.  electron doping, where the two estimates are very close. [{\bf Bottom:}]  The fiduciary  hole doped cobaltate. The two curves in the high $T$ limit  corresponds to the first term in Eq(\ref{s_holedoping}) and from the uncorrelated chemical potential. For the case on right, the Heikes Mott formula misses the enhancement that the high frequency formula predicts. Such enhanced values of the thermopower are very exciting in the current quest for better materials.}
\label{sstar_both}
\end{figure}
\end{flushleft}

We observe that  the first term in Eq(\ref{s1})  from  $\mu(T)$ arising from thermodynamics, termed  the Heikes-Mott contribution,  dominates at very high $T$. The approach to this value is governed by the  second term of Eq(\ref{s1}), called the  transport term. This transport term  is   $O(  \beta t) $ for the triangular lattice, whereas it is only $O(\beta t )^2$ for the square lattice due to the existence of closed loops of length three in the former. The high $T$ expansion  clearly identifies the role of the lattice topology here. The other important consequence is the dependence upon the sign of the hopping in the transport term.  To be specific, for electron doping the thermopower in Eq(\ref{s_electrondoping}) shows that $S$ approaches its high $T$ limit from below as long as $t<0$, as we find for sodium cobaltates\cite{terasaki,ong}. On the other hand, if we could flip the sign of the hopping, as  in a fiduciary hole doped Cobalt Oxide layer,  {\em the high $T$ value would be reached from above}. Since the $S$ must vanish at low $T$, this observation implies that we must find a maximum in $S(T)$ at some intermediate $T$. This then motivates the calculation for a fiduciary system with the flipped sign of hopping. As seen in Fig(\ref{sstar_vs_T},\ref{sstar_both}), numerical results are very encouraging, leading to a thermopower that is $\sim 250 \mu V/K$, and should act as an incentive to the materials community who could seek this type of doping. Crystal structures containing triangular loops are clearly favourable, and this includes several 3D structures as well, such as the FCC and HCP lattices.

\subsection{Lorentz number and Figure of Merit for the  triangular lattice $t$-$J$ model}
We briefly indicate the dependence of the Lorentz number ${\bf L}^*$ and the figure of merit ${\bf Z^*} T$ as computed by us in the case of the triangular lattice \cite{peterson_1,peterson_2}, with parameters appropriate for sodium cobaltates
at $x\sim .68$. Fig(\ref{lzstar}) indicates the dependence of these important parameters on $x,T$ for the $t$-$J$ model clusters of size up to 14. The frequency dependence was estimated to be small and  of the same scale as that of $S(\omega)$, and therefore the results 
are good indicators of the DC values. We must keep in mind, that the finite size effects are  substantial for these small clusters, and hence the behaviour at low $T$ is particularly subject to corrections. Also we stress that our calculation pertains to the electronic 
part of the thermal  conduction, and neglects the often substantial lattice part. Our figure of merit is therefore likely to be a rather optimistic upper bound to the physical values.
Correcting for the lattice part using measured thermal conductivity 
 is straightforward in specific cases.  More non trivially, an elaborate topic treated in recent work, deals with  manipulating the lattice part to maximize the figure of merit by suitably chosen lattice structures and impurities \cite{skutterudites}.
 However, it is clear that several interesting trends emerge from this purely electronic  study.  A striking trend is that the proximity of the Mott Hubbard insulating state $x\sim 0$ is not necessarily favourable for good thermoelectric behaviour with a large $Z^* T$. This is despite the enhancement of $S$ itself. The enhancement arises due to Mott Hubbard correlations that lead to a logarithmic divergence 
of $S$ near half filling\cite{haerter_peterson_shastry_curie_weiss}, but is offset by the unfavourable Lorentz ratio. On the other hand,  the proximity of almost filled bands or almost empty bands seems to be more favourable, and indeed the experience with doping in $Na_xCoO_2$ seems to bear out this finding rather well.
\noindent
\begin{figure}[h]
\includegraphics[width=9cm, angle=-90]{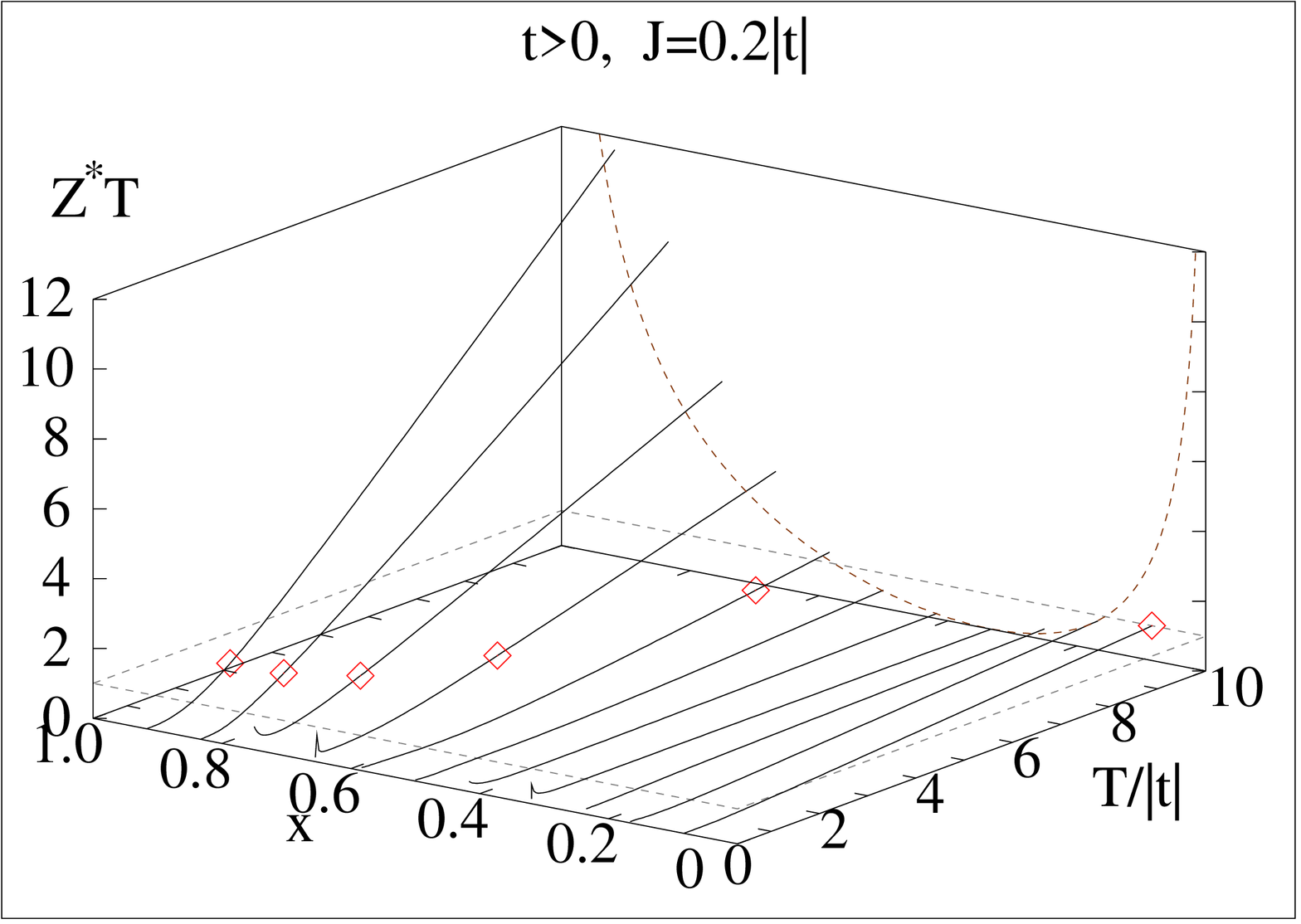}
\caption{[{\bf Left}] The dimensionless figure of merit ${\bf Z}^* T$ versus $x,T$ for the same system from Ref \cite{peterson_2}. This is purely electronic value of the power factor, the phonon contribution to $\kappa$ is expected to be significant and would clearly make the situation  less ``ideal'', thereby we expect the true ${\bf Z}^* T$ to be considerably decreased. However this figure  gives an overview of  the purely electronic contribution to the figure of merit for parameters roughly 
comparable to those in $Na_{.7}CoO_2$. The conclusion in this case is that proximity to half filling is not particularly useful. A similar plot for the fiduciary case of the flipped sign of hopping give a considerably larger  value of the figure of merit.}
\label{lzstar}
\end{figure}

\section{Phenomenological equations for coupled Charge and  Energy transport.}  
In this  section, we  present a simple framework for the problem of  coupled transport of charge and energy (or heat) in a charged  system, such as a semiconductor or a poor metal. It is perturbed by an external temperature gradient and electric fields.  We  add   a  source that dumps energy into the system, such as a  pump laser,  motivated by  several recent experiments\cite{pump_laser_1,pump_laser_2,pump_laser_2.1, pump_laser_3,pump_laser_4,pump_laser_5}. In these  experiments, a  cylindrical rod of a semiconductor (or  metal)  is subjected to  pulsed  laser heating at one end.\footnote{ This end is usually covered by a surface cap of a good metal, where the absorption of the laser power occurs, and is transmitted to the system via a contact layer.}  
The resulting change in the reflectivity of a second probe laser conveniently provides a readout of the local ``dynamical temperature''\cite{pump_laser_1, pump_laser_2,pump_laser_2.1}.  This enables the reconstruction of several physically interesting variables, such as the electron phonon coupling\cite{pump_laser_3}, the thermal conductivity\cite{pump_laser_4} and the thermoelectric coefficient\cite{pump_laser_5}. Given the time dependence  of the probes,  we clearly need  AC response functions of the type discussed above. The laser modulation time constants currently are in the femtosecond range ($10^{-15}$sec), and hence are already able to probe energy relaxation times in semiconductors Ref(\cite{pump_laser_4}).  Such  nontrivial observations  motivate modeling of the type described below.

Our framework  is a simple model of charge and heat  diffusion (e.g. Ref\cite{vollhardt,martin})
together with the exact  coefficients relating the rate of change of currents to the driving forces.
 The dynamical formalism set out in earlier sections is provided a simple context here, and  sheds light on  the meaning of  new operators  $\tau^{xx}, \Theta^{xx}, \Phi^{xx}$  constructed above Eq(\ref{operators}). These  play a fundamental role here, thereby providing a strong pedagogical motivation for this section, in addition to the above mentioned practical one. In our simple model,  we find it advantageous to define new response functions measuring the change in energy, charge density and the currents arising from the input power (the coefficients $M_1, M_2, N_1,N_2$ below in Eq(\ref{new_response_functions})). These are related to the $\vec{q},\omega$ dependent  Onsager coefficients $L_{ij}$ via the various continuity equations and Einstein type relations,  but clearly and neatly isolate the ``force-response'' aspect of the external power  probe, and as such  give a direct method of interpreting experiments.

In essence, we approach this problem through a ``mechanical perturbation'' point of view rather than a thermodynamic one.
Our strategy is to stick to the most broadly definable variables,
such as the energy density and currents, and to avoid, or at least postpone until the very end, mention of variables such as $\vec{q}, \omega$ dependent  {\em temperature fluctuations}. The latter are only sensible in the domain of long wavelength and low frequency variations, unlike the former which are always definable. The results for temperature fluctuations emerge usefully from our formulas in the limiting sense at the end of the calculation.

Let us imagine the system in the form a cylinder of cross section $A$ and length $L$ along the  $x$ axis ($\Omega = L A$), with the surface layer at $x=0$.
 We  subject the system to an external Luttinger field $\psi(\vec{x},t) = \psi_q \exp{- i (\omega t + q_x x)}$, an electrostatic potential $ \phi(\vec{x},t)= \phi_q \exp{- i (\omega t + q_x x)}$. We introduce the pump laser term below via the continuity equation below.
Since we will discuss a charged system, it must be stressed that the electric potential satisfies the Poisson equation with the induced charge density included, so that $\vec{E}= - \nabla \phi $
is the {\em total local electric field}. The system is then described by the Hamiltonian Eq(\ref{hpert1})
We will denote the (Grand canonical) Hamiltonian in the absence of the perturbing fields as $K_0$ and the perturbation as $K_1$. As usual the quantum  average of an observable is given by $\langle Q \rangle = Tr( Q \hat{\rho}(t)) $, where  
the density matrix $\hat{\rho} $ satisfies the von Neumann equation $i \frac{\partial}{\partial t}\hat{\rho}(t) = [K_{tot}(t),\hat{\rho}(t)]$, and hence any observable expectation satisfies the equation $i \frac{\partial}{\partial t} \langle Q(t) \rangle = \langle [Q, K_{1}]\rangle_0 + \langle [Q, K_0]\rangle   $. The first term has been linearized and hence evaluated in the unperturbed ensemble with $K=K_0$, and the second term can be evaluated within perturbation theory as usual, and we find the exact linearized equation of motion:
\begin{equation}
i \frac{\partial}{\partial t} \langle Q(t) \rangle = \langle [Q, K_{1}]\rangle_0 + (-i) \int_0^\infty \, dt'\, e^{ i \omega t' -0^+ t'}  \langle \left[ [Q(t), K_0], K_1(-t') \right] \rangle _0. \;\label{general_EOM}
\end{equation} 
Interestingly enough, the first term in  Eq(\ref{general_EOM}) is expressible exactly in terms of the three operators in Eq(\ref{variables}), while the second term is approximated by a relaxational plus diffusive term. We  choose the variables $Q$ as the heat ($\hat{J}^Q_x$) and charge ($\hat{J}_x$) currents and the densities of heat $K$ and charge $\rho$ as before. Exploiting the linearity of the theory,  it suffices to consider a single frequency and wave vector at the input, and hence  we  introduce a notation for the induced variables depending on the single wave vector $q_x$ through 
\begin{eqnarray}
\langle \hat{J}^Q_x(\vec{x},t) \rangle & = & \frac{1}{\Omega} e^{- i \omega t- i q_x x}  \delta J^Q_x    \nonumber \\
\langle \hat{J}_x(\vec{x},t) \rangle & = & \frac{1}{\Omega} e^{- i \omega t- i q_x x}  \delta J_x    \nonumber \\
\langle K(\vec{x},t) \rangle & = & \frac{1}{\Omega} e^{- i \omega t- i q_x x}  \delta K_q   + \frac{1}{\Omega} \langle K \rangle_0 \nonumber \\
\langle \rho(\vec{x},t) \rangle & = & \frac{1}{\Omega} e^{- i \omega t- i q_x x}  \delta   \rho_q  + q_e n. \label{deltas}
\end{eqnarray}
Thus the variables $\langle \hat{J}^Q_x(\vec{x},t) \rangle $ etc are intensive whereas $\delta J^Q_x$ etc are extensive.  Next, the pump laser  coupling to the system  is introduced via the continuity equation for energy. We  write  the energy continuity equation (ignoring the variations along the transverse directions and focusing on the $x$ axis variation) as 
\begin{equation}
\frac{\partial}{\partial t} \langle K(\vec{x},t) \rangle + \frac{\partial}{\partial x} \langle  \hat{J}^Q_x(\vec{x},t) \rangle  = P_0 \delta(x). \label{continuity} 
\end{equation} 
 We introduced    the input power  $P_0$  per unit area at the surface layer $x=0$.
\footnote {If we introduce the coupling of the laser field $\vec{E}^{0}$ to the matter in $K_{tot}$, it leads to an operator equation of continuity for energy  which contains $\vec{J}.\vec{E}^{0}$ at the surface. Our treatment roughly corresponds to  writing
 $\vec{J}= \sigma_{sur} \vec{E}^{0}$, where $\sigma_{sur}$ is the conductivity at the surface,  and  averaging  this Joule heating over the time period of the laser  $ 2 \pi/\omega_{0}$.  Assume a skin depth $l$ so that the  power absorbed per unit area  $P_0= \frac{1}{2}  l  \sigma_{sur} |\vec{E}^{0}|^2$. Depending on the set up, this might need to be further corrected for a contact (Kaptiza) resistance between the cap and the sample.  }
 The power $P_0$ is further  modulated in time,  so that we decompose it as $P_0 \exp{- i (\omega t)}$. 
 We finally note the conservation laws for charge and energy densities in terms of the response amplitudes :
\begin{eqnarray}
q_x \ \delta J^Q_x+ \omega \ \delta K_q & = &  i A  \ P_0 \nonumber \\
q_x \ \delta J_x + \omega \ \delta \rho_q &=& 0. \label{conservation_laws}
\end{eqnarray}
We next note that the first term in the dynamical equation Eq(\ref{general_EOM}) for both the currents is {\em exactly expressible  in terms of the operators defined in} Eq(\ref{operators}). For example consider the heat current equation where
the term in question reads $\langle [ \hat{J}^Q_x(q), K(-q)] \rangle_0 \psi_q + \langle [ \hat{J}^Q_x(q), \rho(-q)]\rangle_0 \phi_q $. For small $q$, upon using Eq(\ref{operators}) it becomes   $ - \langle \Theta^{xx} \rangle_0 \ q \ \psi_q
- \langle \Phi^{xx} \rangle_0 \ q \ \phi_q$. Similar expressions hold for the charge current. In the absence of the second term in Eq(\ref{general_EOM}), the equations are  ballistic, and hence the sum rules discussed earlier are closely related to the behaviour of the response functions in this regime. This little calculation gives us the physical meaning of these operators $\Theta^{xx}, \Phi^{xx} $; {\em their expectation value   determines the magnitude of the ballistic force exerted by the fields}.  The second term is of course crucial, and we need to estimate it using some general principles. We write the phenomenological equations:
\begin{eqnarray}
\left[ \frac{1}{\tau}+\frac{d}{d t} \right] \langle \hat{J}^Q_x(\vec{x},t) \rangle & = & - \frac{D_Q}{\tau}  \nabla \langle K(\vec{x} t) \rangle  -  \frac{c_1}{\tau}  \nabla \langle \rho(\vec{x} t) \rangle   \nonumber \\ 
&& - \left\{  \frac{ \langle \Theta^{xx}\rangle_0}{\Omega}  \nabla \psi(\vec{x} t) + \frac{\langle \Phi^{xx} \rangle_0}{\Omega}  \nabla \phi(\vec{x} t) \right\} \label{energy_current_EOM}
\end{eqnarray}
and
\begin{eqnarray}
\left[ \frac{1}{\tau}+\frac{d}{d t}\right] \langle \hat{J}_x(\vec{x},t) \rangle & = & - \frac{c_2}{\tau} \nabla \langle K(\vec{x} t) \rangle - \frac{D_c}{\tau}  \nabla \langle \rho(\vec{x} t) \rangle   \nonumber \\ 
&& - \left\{  \frac{ \langle \tau^{xx} \rangle_0}{\Omega}  \nabla \phi(\vec{x} t) + \frac{\langle \Phi^{xx} \rangle_0}{\Omega} \nabla \psi(\vec{x} t) \right\} \label{charge_current_EOM_1}
\end{eqnarray}
These equations represent the effect of the second term in Eq(\ref{general_EOM}) by terms proportional to the gradients of the heat and charge densities, and are relaxational and diffusive in nature. $D_Q, D_c$  are the heat and charge diffusion constants, and the cross diffusion terms $c_1, c_2$ are determined below. The basic physics contained in the diffusion terms is that in steady state  (where the time derivative of the currents are zero), and in the absence of external mechanical fields, one can yet have charge and heat currents driven by gradients of the charge and heat densities.  The relaxation time $\tau\equiv \tau(\vec{q},\omega)$, in general depends upon $\vec{q}, \omega$, and  gives the rate at which the currents relax to zero. Within this model $\tau$   must  necessarily be the same for both charge and energy currents, otherwise the  generalized Onsager reciprocity at finite frequencies  $L_{12}(\omega) = L_{21}(\omega)$ is violated.  Let us also note  the Eq(\ref{charge_current_EOM_1}) in terms of the induced amplitudes in the  Fourier representation: 
\begin{eqnarray}
(1- i \omega \tau) \delta J^Q_x & = & i \tau \left\{  \langle \Theta^{xx}\rangle_0 \ q  \psi_q + \langle \Phi^{xx} \rangle_0  \  q  \phi_q \right\} + i D_Q \ q  \delta K_q + i c_1 \ q  \delta \rho_q  \nonumber \\
(1- i \omega \tau) \delta J_x & = & i \tau \left\{  \langle \Phi^{xx}\rangle_0  \ q  \psi_q + \langle \tau^{xx} \rangle_0  \  q  \phi_q \right\} + i c_2 \ q  \delta K_q + i D_c \ q  \delta \rho_q.  \nonumber \\
\label{charge_current_EOM_2}
\end{eqnarray}
The constants $D_Q, D_c \ ,  c_1, c_2$ can be fixed by considering the static limit at finite $q$ where the currents and their time  derivatives are zero. Equating the various coefficients on the right to zero and taking the long wavelength limit, we determine these as follows:
\begin{equation}
\begin{array}{r l l}
D_Q & =  - \tau \langle \Theta^{xx} \rangle_0  \frac{\psi_q}{\delta K_q}  & =  \tau \frac{\langle \Theta^{xx} \rangle_0}{  C(T) \ T}  \\
D_c  & =  - \tau \langle \tau^{xx} \rangle_0 \frac{\phi_q}{\delta \rho_q}  & =   \tau \frac{\langle \tau^{xx} \rangle_0}{\Omega} \frac{1}{q_e^2} \frac{ d \mu}{ d n}  \\
c_1  & =  D_c \frac{\langle \Phi^{xx} \rangle_0 }{\langle \tau^{xx} \rangle_0},  \hspace{.2in} c_2&   =  D_Q \frac{\langle \Phi^{xx} \rangle_0 }{\langle \Theta^{xx} \rangle_0}  .  
\end{array}\label{diffusion_constants}
\end{equation}
Here $C(T)$ is the (extensive) specific heat, and $ \frac{ d n}{ d \mu}$ is the compressibility per unit volume. We have made use of the standard thermodynamic definitions of these response functions to arrive at these relations. Thus all parameters are fixed in terms of the averages of the three operators Eq(\ref{operators}), a single relaxation time, and the two thermodynamic response functions. For a single frequency mode, these coupled equations together with the conservation laws Eq(\ref{conservation_laws}) can be solved easily,  and the results expressed as 
\begin{equation}
\begin{array}{r l l}
\frac{1}{\Omega} \delta J^Q_x & = & L_{22} ( i q  \psi_q) + L_{21} ( i  q \phi_q) + M_2 \frac{P_0}{L}  \\
\frac{1}{\Omega} \delta J_x& = & L_{12} ( i q \psi_q ) + L_{11}( i q  \phi_q ) + M_1 \frac{P_0}{L}  \label{new_response_functions} 
\end{array}
\end{equation}
In addition to the standard Onsager coefficients $L_{ij}$, we have define   the power response functions $M_j$ as above, giving the response of the currents to $P_0$. It is also interesting to define
the response of the charge and energy density to the applied power $P_0$ via
\begin{equation}
N_1   =  \frac{1}{A} \frac{\partial \delta \rho_q}{\partial P_0},  
  \hspace{.1in} N_2  =  \frac{1}{A} \frac{\partial \delta K_q}{\partial P_0} \label{new_response_functions_2}
\end{equation}
The novel function $N_2$, for example, gives us a measure of the change in energy, and hence temperature, at various points in the system in response to the applied laser heating. We discuss this connection later.

 All of these can be expressed in terms of a convenient energy denominator
\begin{equation}
\Delta = ( 1 - i \omega \tau + i \frac{ D_Q q^2}{\omega})( 1 - i \omega \tau + i \frac{ D_c q^2}{\omega}) + \xi D_c D_Q \frac{q^4}{\omega^2},
\end{equation}
where the dimensionless coupling constant between the charge and heat modes is expressible through the high frequency figure of merit Eq(\ref{hifreq_variables}) as:
\begin{equation}
\xi= \frac{ \langle \Phi^{xx} \rangle _0^2}{\langle \Theta^{xx} \rangle _0\langle \tau^{xx} \rangle _0} =\frac{ Z^* T}{Z^* T + 1}. \label{xi}
\end{equation}
We list the finite $\vec{q}, \omega$  Onsager coefficients:
\begin{eqnarray}
L_{11}& = & \frac{1}{\Delta} \frac{\tau \langle \tau^{xx}\rangle_0}{\Omega} \left[ 1 - i \omega \tau + i (1-\xi) \ D_Q \ \frac{ q^2}{\omega}\right]\nonumber \\
L_{12}& = & \frac{1}{\Delta} \frac{\tau \langle \Phi^{xx}\rangle_0}{\Omega} \left[ 1 - i \omega \tau \right]\nonumber \\
L_{22}& = & \frac{1}{\Delta} \frac{\tau \langle \Theta^{xx}\rangle_0}{\Omega} \left[ 1 - i \omega \tau + i (1-\xi) \ D_c \ \frac{ q^2}{\omega}\right]\nonumber \\
M_{1}& = &  - \frac{1}{\Delta} \ \ D_Q \ \frac{ q    }{\omega }  \  [1- i \omega \tau ] \  \frac{\langle \Phi^{xx}\rangle_0}{\langle \Theta^{xx}\rangle_0} \nonumber  \\
M_{2}& = & - \frac{1}{\Delta} \ D_Q \  \frac{ q \ }{\omega} \left[ 1 - i \omega \tau + i (1- \xi) \ D_c \ \frac{ \ q^2}{\omega}\right]. \nonumber \\
N_1 & = & - \frac{ q}{\omega} M_1  \nonumber \\
N_2 & = & \frac{i}{\omega} - \frac{q}{\omega} M_2.
\label{solutions_1}
\end{eqnarray} 
The coefficient $L_{ij}$ have the standard meanings that we have commented upon earlier. The coefficients $M_1, M_2, N_2$ etc are the response coefficients to the applied power source. It is easy to see that these dynamical results satisfy the sum rules in Eq(\ref{plasma_sumrule}, \ref{thermal_sumrule_1},\ref{thermal_sumrule_2}).

 We note several  points about this exercise next.
\begin{enumerate}
\item The above expressions are  written in terms of energy variables. It  is more rigorous as well as  profitable to view the transport processes as primarily those of charge and { energy}, rather than temperature. While we can always define an energy fluctuation, it translates to a temperature pulse only under conditions of local equilibrium, which might not always be attainable. 
\item These coupled equations have a similarity to those in the description of the two coupled fluids in  $^4He$\cite{landau_lifshitz}.  For insulators, the coupling between the lattice energy modes and lattice displacement modes\cite{martin, guyer} also has a formally similar structure. In these cases, the role of our coupling  parameter $\xi$ is played by the dimensionless constant $C_p/C_v -1$. 
\item The new response functions $M_j, N_j$ shed some light on pulse probe type experiments. The coefficient $N_2$ Eq(\ref{new_response_functions_2}) is of particular interest.    In an experiment with  pulsed laser heating, one would use the coefficient $N_2$ to compute the induced energy change $\delta K_q$. This  fluctuation  is interpretable as a temperature variation only if the frequency is low enough to achieve local equilibrium, but is always definable as a mechanical response function. With the help of a model of the above type, it can be used to extract the diffusion constant and hence the thermal conductivity. 
\item  For illustration of the above comment, if we turn off the coupling between the charge and energy modes (set $\xi=0$) then $N_2 \rightarrow i(1- i \omega \tau)/( \omega - i \omega^2 \tau + i D_Q q^2)$. By cross multiplying we can rewrite this in the  suggestive form $\left[ \omega + i \frac{ D_Q q^2}{1- i \omega \tau}\right]\delta K_q = i P_0 A $. If we take the limit of a slow response then we may express the fluctuation in terms of the temperature fluctuation   $\delta K_q = C(T) \delta T_q$. This can  seen to be of the form proposed  by Cattaneo\cite{cattaneo} as an improvement of Fourier's law
\footnote{ The standard argument for  the Cattaneo equation is that   Fourier's law is replaced by $(1 + \tau \frac{\partial}{\partial t}) J^Q_x = \Omega \kappa(0) (-\nabla T)$, where $\kappa(0)$ is the DC thermal conductivity. Combining with the energy conservation law Eq(\ref{conservation_laws}) and further writing all variations in terms of those of the temperature $\delta T_q$ as $\delta K_q = C(T) \delta T_q$, we find $C(T) \left[ \omega + \frac{i \kappa(0) \Omega q^2}{C(T)(1-i \omega \tau)} \right] = i A P_{0}$, which is the same as the previous result on using the relation $\kappa(0)= D_Q C(T)/\Omega$.}.
\item In the decoupled limit $\xi=0$, we find  $L_{22}= \frac{\tau \langle \Theta^{xx} \rangle_0 }{\Omega(1-i\omega \tau + i D_Q q^2 \omega)}.$
The form of $L_{22}$ displays the possibility of a propagating mode for $\omega \tau \gg 1$, with a dispersion $\omega \sim |q| \sqrt{D_Q/\tau}$, corresponding to a second sound. The velocity of the second sound in this simplified model is also expressible in terms of the average of the $\Theta^{xx}$ operator from Eq(\ref{diffusion_constants}). 
\item We see from this framework that one may devise  experiments to isolate and measure different 
terms in the response functions. In particular, the thermal sum rule Eq(\ref{thermal_sumrule_1}) is given in terms of the expectation of  $\langle \Theta^{xx} \rangle$, and one might ask how this can be measured. In response to a $\delta(t)$ pulse of temperature,  the induced heat current pulse in the time domain  jumps at $t=0$ and the magnitude of the jump is free from $\tau$ and a function of $\langle \Theta^{xx} \rangle$ only. Similarly the energy density (hence the local reflectivity) contains an initial $t \theta(t)$ linear rise  \footnote{In the decoupled limit, the energy density satisfies an equation (in space time variables) $\frac{\partial^2}{\partial t^2} K(\vec{x},t)= \frac{D}{\tau} \frac{\partial^2}{\partial x^2} K(\vec{x},t) - \frac{1}{\tau}  \frac{\partial}{\partial t} K(r,t) + \frac{\langle \Theta^{xx} \rangle}{\Omega} (- \nabla^2 \psi)$. From this we see that a $\delta(t)$ pulse in a spatially varying $\psi$ would give an initial linear rise in the energy $\sim t \theta(t)$, with a slope that is inertial, i.e.  independent of $\tau$.}
\item In the decoupled limit,  the electrical conductivity $ \sigma_{xx}(\omega) = \frac{\sigma_0}{1- i \omega \tau + i D_c q^2/\omega},$ with 
$\sigma_0= \tau \langle \tau^{xx} \rangle_0/\Omega$. Hence the dielectric function

$\epsilon(q,\omega) = 1 + \frac{ 4 \pi i }{\omega} \sigma $, has the correct limiting forms for a metal in both static and the plasmon limits. In the static limit we find the screening behaviour $\epsilon = 1 + \frac{4 \pi q_e^2}{q^2}  \frac{d n}{ d \mu}$. For large frequencies $\omega \tau \gg 1$ we get the plasmon behaviour $\epsilon= 1- \frac{\omega_p^2}{\omega^2}$ with $\omega_p^2= \frac{4 \pi}{\Omega} \langle \tau^{xx} \rangle_0$.
\item For a dense metallic system we are usually in the limit where the energy scales are such that $ q v_F \gg \omega$, so that it is a good approximation to regard the charge redistribution as almost instantaneous compared to the heat diffusion. More formally in the decoupled limit we can see that the current response can be written in  suggestive alternate forms
\begin{eqnarray}
\delta J_q & = & \frac{ \tau \langle \tau^{xx} \rangle_0 }{1- i \omega \tau + i \frac{D_c q^2}{\omega}}\ i q  [  \phi_q] \nonumber \\
&=& \frac{ \tau \langle \tau^{xx} \rangle_0 }{1- i \omega \tau }\ i q \ [  \phi_q + \frac{1}{q_e^2 \Omega} \frac{ d \mu}{ d n} \delta \rho_q ].
\end{eqnarray}
We have used the conservation law to go from the first form to the second. If we assume very fast relaxation of the charges, then the term in square brackets can be rewritten approximating the $\delta \rho_q$ term by its static counterpart, resulting  in  the familiar {\em electrochemical potential} $- \nabla (\phi + \frac{1}{q_e} \mu[ n(r)])$. However, for poorly screened low density electron systems and for narrow band systems, it is better to avoid this approximation.
\item These dynamical expressions have the property that the Seebeck coefficient and the Lorentz number are frequency independent, and hence the high frequency approximation Eq(\ref{hifreq_variables}) is exact here.
\end{enumerate}  

Thus we see that these simple equations illustrate the meaning and possible applications of energy transport in novel situations. We can  make contact with standard transport theory in the limit of slow long wavelength variations, where  we have argued that the Luttinger field is equivalent to a temperature field through $\nabla \psi(r)=  \nabla T(r) /T $.

\section{Conclusions}
In this article we have presented the basic ideas of a novel approach to computing certain interesting transport coefficients for correlated systems. These include the important  Seebeck coefficient and the figure of merit. Our basic formalism  extends the idea first used by Luttinger, namely that a gravitational field can be used as a mechanical proxy for temperature gradients.  We take this view point further to include arbitrary time variations, thereby  enabling an exploration of the  regions of frequency  that are normally precluded  in dealing with temperature variations. This   leads to the recognition of a new sum rule for thermal conductivity, as well the application of high frequency ideas to compute the response functions mentioned above. We describe  quantitative applications of these ideas in the context of the properties of the recently found sodium cobaltate materials.

 A well defined program to correct the high frequency results for the effects of finite  frequency, and hence to approach the transport limit is formulated and illustrated.  A  simple phenomenological framework with the novel response functions $M_j, N_j$ is  described, where the role of the newly defined operators becomes clear. This framework,  and its many possible extensions to include other densities, should be of useful in formulating   the new class of experiments made possible by pulsed laser heating.
  
{\bf Acknowledgment}: This work was supported by    grant  NSF-DMR 0706128, and by  grant DOE-BES DE-FG02-06ER46319. It is a pleasure to thank  M. Peterson, 
 S. Mukerjee,  Y. Ezzahri and  A. Shakouri  for helpful  discussions.  I  thank   N. P. Ong for providing me with experimental inputs that have been invaluable. It is a pleasure to acknowledge stimulating comments from  P. W. Anderson regarding the Kelvin-Onsager debate.



\begin{thebibliography}{99}
\bibitem{lectures}  Based on lectures given at the  43rd Karpacz Winter School of Theoretical Physics.  Lecture notes:    {\em Condensed Matter Physics in the Prime of XXI Century:}, Ed. J. Jedrzejewski,(World Scientific, New York 2008).
\bibitem{reviews_1} P. W. Anderson, P. A. Lee, M. Randeria, T. M. Rice, N. Trivedi and F. C. Zhang, J. Phys. Condens. Matter {\bf 16} R755-R769 (2004). 
\bibitem{reviews_2} J. Orenstein and     A. J. Millis, Science, {\bf 288}   468  (2000).  
\bibitem{reviews_3} J. Merino  and R. H. McKenzie, Phys. Rev. {\bf B 61} 7996 (2000). 
\bibitem{reviews_4}  O. Fischer , M. Kugler, I. Maggio-Aprile, and C. Berthod, Rev. Mod. Phys. {\bf 79} 353 \textsc{}(2007).
\bibitem{lee} P. A. Lee, N. Nagaosa and X. G. Wen, Rev. Mod. Phys. {\bf 78 }, 17 (2006); X. G. Wen and P. A. Lee, Phys. Rev. Letts. {\bf 76 }, 503 (1996)
\bibitem{fermi_liquid_theory_anomalies_1} M. Cyrot, Nature {\bf 330}, 115 (1987); H. Fukuyama, Jour. Phys. Chem. Solids {\bf 59}, 447 (1998).
\bibitem{hubbard_tj} A. B. Harris and R. V. Lange, Phys. Rev. {\bf 157} 295 (1967); B. Hetenyi and H. G. Evertz, ArXiv:0806.3848(2008).
\bibitem{ashcroft_mermin} N. Ashcroft and N. D. Mermin, {\em Solid State Physics}, (Harcourt Brace Jovanovich College Publishers,  Fort Worth, 1976).
\bibitem{mahan}G. D. Mahan, {\em Many Particle Systems} ( Plenum, New York, 1990).
\bibitem{fetter}  A. L. Fetter and J. D. Walecka, {\em Theory of Many Particle Systems} (Dover, New York 2003).
\bibitem{sommerfield}  A. Sommerfeld and N. H. Frank,  Rev. Mod. Phys.{\bf 3}, 1 (1931).
\bibitem{bloch} F. Bloch, Zeit.  Phys. {\bf 52}, 555 (1928), ibid {\bf 59}, 208 (1930).
\bibitem{wilson} A. H. Wilson, Proc. Roy. Soc. {\bf A 133}, 458 (1931).
\bibitem{fermi_liquid_theory} Philippe Nozier\`es, {\em Theory of Interacting Fermi Systems}, (Westview Press 1997, New York).
\bibitem{mott_insulator} J H de Boer and E J W Verwey, Proc.  Phys. Soc.  London {\bf 49}, 59 (1937);  N F Mott and R Peierls, Proc.  Phys. Soc.  London {\bf 49}, 72 (1937);  N F Mott, Proc.  Phys. Soc.  London {\bf 62 } , 416 (1949).
\bibitem{pwa_early} P. W. Anderson, Nature {\bf 235}, 1196(1987).
 \bibitem{fermi_liquid_theory_anomalies_2}  C. M. Varma, P. B. Littlewood, S. Schmitt-Rink, E. Abrahams, and A. E. Ruckenstein, Phys. Rev. Lett. {\bf 63}, 1996 (1989).
\bibitem{terasaki} I. Terasaki, Y. Sasago and K. Uchinokura, Phys. Rev. {\bf B 56}, R12685 (1997).
\bibitem{ong} Y. Wang, N. S. Rogado,  R. J. Cava and  N. P. Ong, Nature {\bf 423}, 425 (2003).
\bibitem{mahan_review} G. D. Mahan, Solid State Phys. {\bf 51}, 81(1998),   M. V. Simkin and G. D. Mahan Phys. Rev. Lett. {\bf 84}, 927  (2000).
\bibitem{dmft} G. Kotliar and D. Vollhardt, Physics Today {\bf 57}, 53 (2004); A. Georges, G. Kotliar, W. Krauth, and M. J. Rozenberg Rev. Mod. Phys. 68, 13 (1996).
\bibitem{kotliar}  G. Palsson and G. Kotliar, Phys. Rev. Lett. 80, 4775 (1998), V. S. Oudovenko and G. Kotliar, Phys. Rev. B 65, 075102 (2002); Th. Pruschke,  M. Jarrell, J. K. Freericks, Adv. Phys. {\bf 44} 187 (1995).
\bibitem{zlatic}   V. Zlatic and R. Monnier Phys. Rev. {\bf B 71}, 165109 (2005), J. K. Freericks, D. O. Demchenko, A. V. Joura, and V. Zlatic,
Phys. Rev. {\bf B 68}, 195120 (2003), V. Zlatic, B. Horvatic, I. Milat, B. Coqblin, G. Czycholl, and C. Grenzebach Phys. Rev. {\bf B 68} , 104432 (2003). 
\bibitem{zlatic_rmp}   J. K. Freericks and V. Zlatic, Rev. Mod. Phys. {\bf 75}, 1333 (2003).
\bibitem{ssshall}  B. S. Shastry, B. I. Shraiman and R. R. P. Singh, Phys. Rev. Lett. {\bf 70}, 2004 (1993).
\bibitem{ziman} J M Ziman, {\em Principles of the Theory of Solids}, (Cambridge University Press, Cambridge, 1969).
\bibitem{argyres} P. N. Argyres and J. L. Sigel, Phys. Rev. {\bf B 9}, 3197 (1974); H. Maebashi and H. Fukuyama, Jour. Phys. Soc. Japan {\bf 66}, 3577 (1997). 
\bibitem{peierls} R. E. Peierls, Ann. Physik {\bf 3}, 1055 (1929).
\bibitem{haerter_shastry_hall_2007} J. O. Haerter and B. S. Shastry,  arXiv:0708.1651, Phys. Rev. {\bf B 77}, 045127 (2008).
\bibitem{kumar_shastry}B. Kumar and B. S. Shastry, Phys. Rev. {\bf B 69}, 059901 (2004),  (E) Phys. Rev. {\bf B 68}, 104508 (2003).
\bibitem{imada} F.F. Assad and M. Imada, Phys. Rev. Lett. {\bf 74}, 3868 (1995)
\bibitem{drew}A. T. Zheleznyak, V. M. Yakovenko, and H. D. Drew, Phys. Rev. {\bf B 57}, 3089 (1998).

\bibitem{holstein} T. Holstein, Phys. Rev. {\bf 124}, 1329 (1961), Phil. Mag. {\bf 27}, 225 (1973).
\bibitem{wang_ong}Y. Wang, N. S. Rogado, R.J. Cava and N. P. Ong, arXiv:cond-mat/0305455.  For a smaller range of temperatures the same T linear behaviour is shown in Fig.(3.c) of
M. L. Foo, Y. Wang, S. Watauchi, H.W. Zandbergen, T. He, R. J. Cava, and N. P. Ong,  Phys. Rev. Letts. {\bf 24 }, 247001 (2004). See also E. J. Choi, S. H. Jung, J. H. Noh, A. Zimmers, D. Schmadel, H. D. Drew, Y. K. Bang, J. Y. Son, and J. H. Cho, Phys. Rev. {\bf 76 }, 033105 (2007). 
\bibitem{haerter_peterson_shastry_curie_weiss}
J. O. Haerter, M. R. Peterson, and B. S. Shastry, Phys. Rev. Lett. {\bf 97}, 226402 (2006), Phys. Rev. {\bf B 74}, 245118 (2006)
\bibitem{singh} D. J. Singh, Phys. Rev. {\bf B 61 }, 13397 (2000); A.O. Shorikov and V.I. Anisimov and M.M. Korshunov, ArXiv.org/0705.1408 (2007).
\bibitem{shastry_sumrule} B. S. Shastry, Phys. Rev. {\bf B 73} 085117(2006); (E) Phys. Rev. {\bf B 74} 039901 (2006). An updated version with several errors removed is available at \url{http://physics.ucsc.edu/~sriram/papers_all/ksumrules_errors_etc/evolving.pdf}. 
\bibitem{onsager} L. Onsager, Phys. Rev. {\bf 37}, 405 (1931), Phys. Rev. {\bf 38}, 2265 (1931).
\bibitem{luttinger} J. M. Luttinger, Phys. Rev. {\bf 135}, A 1505 (1964), Phys.Rev.
{\bf 136 }, A1481 (1964). 
\bibitem{compressibility_sum_rule} J. M. Luttinger and P. Nozi\`eres, Phys. Rev. {\bf 127}, 1423, 1431 (1962).
\bibitem{kelvin} W. Thomson (Lord Kelvin)  Proc. Roy. Soc. Edinburgh: p.123 (1854),Collected Papers I, pp. 237-41. 
\bibitem{kubo} R. Kubo, J. Phys. Soc. Japan {\bf 12},570 (1957).
\bibitem{fpu} E. Fermi, J. Pasta, S. Ulam, Studies of Nonlinear Problems, Document Los Alamos-1940, Unpublished (May 1955).
\bibitem{fsumrule}R. Bari, D. Adler and R. V. Lang,   Phys. Rev. {\bf B 2}, 2898(1970);  E.Sadakata and E Hanamura, J Phys. Soc. Japan {\bf 34}, 882 (1973); P. F. Maldague, Phys. Rev. {\bf B 16}, 2437 (1977).
\bibitem{beni} P. Chaikin and G. Beni, Phys. Rev. {\bf  B 13}, 647 (1976).
\bibitem{heikes}  R. R.  Heikes, {\em Thermoelectricity} (Wiley-Interscience, New York, 1961)
\bibitem{wannier} G. H. Wannier, {Statistical Physics}, pp506 (Dover Publications, NY 1966).
\bibitem{hubbardops} The Hubbard operators are defined in J. Hubbard, Proc. Roy. Soc. {\bf A 276},  238(1963);
{\bf A 277},  237 (1964); {\bf A 281},  401 (1964); {\bf A 285},  542 (1964).
\bibitem{peterson_1} M. Peterson, S. Mukerjee,  B. S. Shastry and J. O. Haerter, Phys. Rev. {\bf B 76 }, 
125110 (2007).
\bibitem{peterson_2} M. Peterson,   B. S. Shastry and J. O. Haerter arXiv:0705.3867;  Phys. Rev. {\bf B 76}, 165118 (2007).
\bibitem{skutterudites} G. S. Nolas, D. T. Morelli, and T. M. Tritt, Annu. Rev. Mater. Sci. {\bf 29}, 89 (1999).N. R. Dilley {\it et. al.} Phys. Rev. {\bf B 61}, 4608 (2000);
\bibitem{pump_laser_1} C. A. Paddock and G. L. Eesley, J. Appl. Phys. {\bf 60}, 285 (1986), D. Cahill,  Rev.  Sci.  Instr. {\bf 75}, 5119 (2004).
\bibitem{pump_laser_2} J. G. Fujimoto, J. M. Liu, E. P. Ippen and N. Bloembergen, Phys. Rev. Letts. {\bf 53}, 1837 (1984).
\bibitem{pump_laser_2.1} D. G. Cahill,  Rev.  Sci. Instr.  {\bf 75}, 5119 (2004).
\bibitem{pump_laser_3} P. B. Allen, Phys. Rev. Letts. {\bf 59}, 1460 (1987).
\bibitem{pump_laser_4} Y. K. Koh and D. G. Cahill, Phys. Rev. {\bf B 76}, 075207, 2007.
\bibitem{pump_laser_5} S. Zeuner, H. Lengfellner and W. Prettl, Phys. Rev{\bf B 51}, 11903 (1995).
\bibitem{vollhardt} D. Vollhardt and P. Wolfle, Phys. Rev. Lett. {\bf 45}, 842 (1980).
\bibitem{martin} P. C. Kwok and P. Martin, Phys. Rev. {\bf 142}, 495 (1966) .
\bibitem{landau_lifshitz} L. Landau and E. Lifshitz, {\em Fluid Mechanics}, Second Edition: Volume 6 (Course of Theoretical Physics),  Butterworth-Heinemann, Oxford (2000).  
\bibitem{guyer} R. A. Guyer and J. Krumhansl,  Phys. Rev {\bf B 148}, 778 (1966).
\bibitem{cattaneo} D. D. Joseph and Luigi Preziosi, Rev. Mod. Phys. {\bf 61}, 41 (1989).
\end{thebibliography}
\end{document}